\documentclass[a4paper,10pt]{article}
\usepackage{amssymb}
\usepackage{amsfonts}
\usepackage{amsmath}
\usepackage{graphicx}
\input epsf    %  carica MACRO eps.tex (WinEdt sa dov'e'.....)

\newtheorem{definition}{Definition}[section]
\newtheorem{theorem}{Theorem}[section]

\newtheorem{corollary}[theorem]{Corollary}

\def\à{\`a}
\def\ò{\`o}
\def\ì{\`\i}
\def\ù{\`u}
\def\à{\`a}
\def\è{\`e}
\def\é{\'e}
\def\È{\`E}

\def\dlangle{\langle\!\langle}
\def\drangle{\rangle\!\rangle}
\def\dvert{\parallel}

\begin{document}

\title{\bf On von Neumann's Examples of Types}

\author{{\small Renato Nobili -- Dipartimento di Fisica ''G.
Galilei'', Universit\à di Padova,}\\{\small via Marzolo  8,  35131 Padova
--- ITALY} }

%\vspace{1truecm}
\date{\small 10 September 2008}
%\vspace{1truecm}
\maketitle

\begin{abstract}
{\small \noindent The factorization properties of operator algebras in
separable Hilbert spaces was developed by John von Neumann in collaboration
with F.J. Murray in four outstanding papers published from 1936 to 1943.
Unfortunately, probably because of conceptual difficulties with physical
interpretations, some relevant results presented in those papers, in
particular the remarkable examples of factors, have not been adequately
considered so far. The paper here presented aims to introduce the subject in a
new although maybe unusual form pursuing three main goals: speculating about
the physical reasons and motivations that are likely to have been at the
origin of von Neumann's investigation; describing the examples of factors
provided by von Neumann and Murray with the purpose of clarifying the general
concepts standing at the base of the classification of algebraic factors into
three general types; outlining the perspective of extending the theory to
non--separable Hilbert spaces with the purpose of suggesting a novel approach
to the representation of infinite systems controlled by external gauge fields.}
\end{abstract}

\quad

\section*{Introduction}
\noindent According to the {\em correspondence principle} stated by Niels Bohr
(1923), the behavior of a quantum mechanical system approaches that of a
classical system in the limit of large quantum numbers. As shown by Dirac in
1926, this principle can be mathematically formulated by saying that Poisson's
brackets are the classical limits of Heisenberg's commutators divided by
$i\hbar$. Correspondingly, the transformations $q' =T_c\, q$, $p' = T_c\, p$ of
the  classical  phase--space variables $\{q, p\}$ are found to be the classical
limits of the algebraic automorphisms $Q'=TQT^{-1}$, $P' = T P T^{-1}$ acting
on the operators $\{Q,P\}$, which play the role of $\{q, p\}$ in the
quantum--mechanical representation\footnote{This is in contrast with the
attempts to unify the theory of the classical and the quantum--mechanical
systems basing on the correspondence between physical states rather than
between classical transformations and observable automorphisms.}.

In other words, the Abelian groups of classical transformations appear to be
isomorphic to the Abelian groups of quantum--mechanical automorphisms, rather
than to the group of transformations $\vert \phi'\rangle = T\vert \phi\rangle$
acting on the unit vectors $\vert \phi\rangle$ ({\em rays}), which represent
the states of the quantum--mechanical system. The latter, indeed, are
non--Abelian projective extensions of the Abelian groups of phase--space
transformations $T_c$ \cite{LEVY}.

So, while in classical mechanics the algebraic variables that represent the
physical quantities differ qualitatively from the differential operators that
represent the generators of physical transformations, in quantum mechanics, by
contrast, the Hermitian operators that represent the physical quantities
represent also the generators of physical transformations; as if {\em
observing} and {\em operating} were two complementary aspects of physical
phenomenology.

A few years later, in {\em The theory of groups and quantum mechanics, \S 14},
\cite{WEYL} (translation of the German edition published in 1929), Hermann Weyl
declared to feel certain that the {\em kinematical structure of a physical
system is expressed by an irreducible Abelian group of unitary ray rotations
in system space} (which is a different name for "irreducible non--Abelian
projective extension of an Abelian group"). This structure leads automatically
to the decomposition of a family of independent observables into pairs of
conjugated quantities. For quantum--mechanical systems with classical analog,
the group is generated by the observables $Q$ e $P$. However, as proved by
Weyl himself, also the commutative and anti--commutative algebras generated by
the creation and destruction operators of bosonic and fermionic fields can be
derived from Abelian groups of automorphisms.

In this view, the general structure of quantum--mechanical systems emerges as
a maximal commutative algebra $\cal P$ of projectors $P, P', \dots$,
representing a complete set of compatible observables, which are equipped with
an Abelian group $\cal U$ of automorphisms $P' = U P U^{-1}$ induced by a
group of unitary operators $U\in \cal U$, which can be interpreted as the
carriers of canonical--coordinate transformations. It is worth noticing,
however, that after the discovery of the internal--symmetry groups of
elementary particles there is no reason to require that the automorphism
groups devised by Weyl be Abelian.

The fact that the transformations acting on the states of a classical system
are one--to--one with the algebraic automorphisms acting on the observables of
a corresponding quantum--mechanical system admits the following physical
interpretation: {\em both groups represent the operational possibilities of an
ideal macroscopic observer interacting with the system}. It is therefore quite
natural to assume that, since these possibilities are given in the macroscopic
world, they can be represented as a set of {\em mutually exclusive events}, as
they were the possible results of macroscopic measurements. It is therefore
reasonable to assume that a quantum--mechanical representation of this set of
operational possibilities be implemented by labeling the vectors of a
Hilbert--space basis by the elements of a suitable group $G$. In this way, the
observed--observer system could find a suitable representation in the direct
product of the Hilbert space formed by the state vectors of the observe system
with that spanned by a basis of vectors labeled by the elements of $G$.

An evident disadvantage of this description is that, if we require that the
Hilbert space of operational possibilities is separable, then only discrete
groups of transformations can be admitted. {\em En passand}, this leads to
hypothesize that the description of a more realistic set of operational
possibilities, possibly parameterized by external gauge fields, could be given
only in non--separable Hilbert spaces.

By the arguments that we are just on the point to introduce, we want to prove
that the direct product of two Hilbert spaces of the type described above
produces a {\em hybrid space}, in which the algebraic factorization of the
entire system in two independent, although mutually correlated parts, has its
most natural representation.

Such a factorization is not arbitrary but conditioned by the algebraic
structure of the observables of the composed system. Throughout this study, we
will elucidate also the problem of classifying and representing the ways in
which the parts of a quantum--mechanical system interact with each other. The
relevance of this fact lies in that {\em this problem does not find a
satisfactory solution if the state space of the composed system is naively
assumed as the direct product of the state spaces of the parts}.

For instance, if the state of a system formed by two parts ${A},{A}'$ is
represented as a unit vector $\vert\Phi\rangle$ belonging to the direct product
${\cal H}$ of the Hilbert spaces ${\cal H}_A$, ${\cal H}_{A'}$ of the parts,
we have
$$\vert\Phi\rangle = \sum_{ij} c_{ij}\vert\alpha_i\rangle \otimes \vert\alpha'_j\rangle\,,$$ where
$c_{ij}$ are complex constants and $\vert\alpha_i\rangle$,
$\vert\alpha'_j\rangle$ are the basis vectors respectively of ${\cal H}_A$ and
${\cal H}_{A'}$.

In this way, the composition of the two parts produces the {\em entanglement}
of their respective states. This has no analog in classical mechanics. Now, it
is known that, by keeping $\vert\Phi\rangle$ fixed, the bases of ${\cal H}_A$,
${\cal H}_{A'}$ can be rotated so as to obtain the one--index summation
$$\vert\Phi\rangle = \sum_{i} \sqrt{w_i}\vert\alpha_i\rangle \otimes
\vert\alpha'_i\rangle \,,$$ where the positive constants $w_i$ can be
interpreted as the probabilities that a suitable observation of $A$, or of
$A'$, detects that the state of the system is $\vert\alpha_i\rangle \otimes
\vert\alpha'_i\rangle$. We immediately see how the entanglement of part states
results into a discrete pairing of part states. Of course, if
$\vert\Phi\rangle$ varies in ${\cal H}$, also the states of each pairing and
the values of $w_i$ in general change.

Compositions and measurements of this sort explain how, for instance, the
observation that a measurement device $A'$ is in a state
$\vert\alpha'_i\rangle$ determines the simultaneous projection of the state of
the measured system $A$ into the state $\vert\alpha_i\rangle$ paired with
$\vert\alpha'_i\rangle$ in $\vert\Phi\rangle$.

Clearly, these sorts of pairings are possible only if the quantities to be
measured possess discrete sets of eigenvalues. But how then could we deal with
quantities whose eigenvalue--spectra are with continuous or with infinite
collections of quantities?

The problem of measurement and observation of quantum--mechanical systems was
carefully analyzed by von Neumann. In his famous book of 1932 on the
Mathematical Foundations of Quantum Mechanics \cite{NEU}, after discussing the
process of measurement as outlined above, the great author posed also the
problem of how, starting from an initial state of the form $\vert\Phi_0\rangle
= \vert\phi_0\rangle\otimes \vert\phi'_0\rangle$, i.e., a state formed by two
non--entangled states $\vert\phi_0\rangle$ and $\vert\phi'_0\rangle$, the
entangled state $\vert\Phi\rangle = \sum_{i} \sqrt{w_i}\vert\alpha_i\rangle
\otimes \vert\alpha'_i\rangle$ can be generated.

To capture the problem in its very essence, von Neumann assumed that the states
of the measurement device, and consequently also those of the observed system,
be indexed by the elements of a discrete group, precisely the Abelian group of
integers $Z$.

In particular, he assumed that the initial resting--state  of the measurement
device is $\vert\phi_0\rangle= \vert\alpha'_0\rangle$ and that the initial
state of the system under observation is $$\vert\phi_0\rangle= \sum_{n\in Z}
\sqrt{w_{n}}\vert\alpha_n\rangle\,;$$ that is, a superposition of the
eigenstates of the quantity under observation, whose probability amplitudes
$w_n$ can be related to the statistical frequencies of results obtained in a
large number of measurements repeated in identical conditions. The question is
then to explain how the transition
$$
\vert\Phi_0\rangle = \Bigl(\sum_{n\in Z} \sqrt{w_n}\vert\alpha_n\rangle\Bigr)
\otimes \vert\alpha'_0\rangle \quad\Longrightarrow\quad \vert\Phi\rangle =
\sum_{n\in Z} \sqrt{w_n}\Bigl(\vert\alpha_n\rangle \otimes
\vert\alpha'_n\rangle\Bigr)\,,
$$
from the initially disentangled state $\vert\Phi_0\rangle$ to the entangled
state $\vert\Phi\rangle $ can occur before the observation of the measurement
device disentangles the pair $\vert\alpha_n\rangle \otimes
\vert\alpha'_n\rangle$ with probability $w_n$.

Von Neumann's answer was that the transition $\vert\Phi_0\rangle
\Rightarrow\vert\Phi\rangle$ can be interpreted as a temporal evolution
$\vert\Phi\rangle = U\,\vert\Phi_0\rangle$ performed by the unitary operator
defined by the equations $$U \vert\alpha_n\rangle \otimes \vert\alpha'_m\rangle
= \vert\alpha_n\rangle \otimes \vert\alpha'_{n+m}\rangle\,.$$ This operator,
which unfortunately cannot be expressed as a continuous function of time
because of the discreteness of state entanglement, has the form
$$
U = \sum_{n\in Z} P_n\otimes T^n\,,\quad P_n =
\vert\alpha_n\rangle\langle\alpha_n\vert\,,\quad T= \sum_{n\in
Z}\vert\alpha'_{n+1}\rangle\langle\alpha'_{n}\vert\,.
$$

Thus, ultimately, the problem of how state entanglement may take place during
observation processes leads naturally to represent the observed--observer
system in the framework of a hybrid space ${\cal H}= {\cal H}_X\otimes {\cal
H}_Z$, where $X$ is the index set of the state--projectors $P_n$ family of the
observed system and $Z$ the translation group of the observing--device pointer.

The celebrated papers of von Neumann (and Murray) on the rings of operators,
\cite{NEU1} \cite{NEU2} \cite{NEU3} \cite{NEU4}, seem to be directed to
exhaust the analysis of the measurement problem, which was introduced only
schematically and as an example in the quoted book on Foundations. Those
papers seem to indicate that, at the maximum level of generality,  the
solution to the problem can be reached by pursuing the goal of understanding,
not just how the states of the parts can pair with each other, but, more
basically, how the operators that represents the observables of the parts can
correlate with each other in a measurement process.

In the following, we will see how the operator algebras that can be formed
quite naturally in the hybrid spaces of the sort described above, afford a
wonderful territory in which the investigation of this subject can be carried
out in the most fruitful way.

We will limit ourselves to the case, completely solved by von Neumann, in which
the Hilbert space of operational possibilities is separable.  That is, the case
in which the operations form a non--Abelian discrete group $G$. In the most
natural way, and with maximal generality, the structure of the observed system
will be represented by an algebra of projectors indexed by the measurable
subsets (in the most general sense) of a set $X$, in which $G$ acts by point
transformations. For the sake of language simplicity, systems of this sort
will be called {\em discrete}. For reasons that will be clear in a second
moment, we will assume that $G$ acts on $X$ {\em freely}, i.e., with no fixed
subsets of non--zero measure, and {\em ergodically}, i.e., transitively for
non--zero--measure subsets of $X$.

This treatment differs somewhat from that of the great Austro--Hungarian
mathematician and of his coworker. In certain sense, it is a considerable
simplification of their original treatment (besides, of course, an incomplete
and lacunal exposition of it). Nevertheless, we hope that by pursuing the goal
of evidencing the physical meanings of the construction and by using a
formalism based on the definition of certain operators in the space of
operational possibilities, we will be able (hopefully) to make the subject
simpler and more transparent.

Before entering the subject, we will introduce without any pretensions of
completeness and precision some basic concepts regarding the algebras of
observables.

\section{The theory of factorization}
We will summarize here the most important results obtained by von Neumann on
the algebras of bounded operators in Hilbert spaces (the algebras of unbounded
operators exhibit incurable pathologies). In the following, the term {\em
algebra} will indicate a $C^*$--algebra, i.e., an algebra containing the
adjoint $A^\dag$ of every operator $A$.

\subsection{Von Neumann algebras}
\begin{definition}
Let ${\cal A}$ be an algebra of bounded  operators of the Hilbert space $\cal
H$. Precisely, a subalgebra of the algebra ${\cal I}$ formed by all the bounded
operators of $\cal H$. Call the {\em commutant} of ${\cal A}$ the algebra
${\cal A}'$ formed by all bounded operators of $\cal H$ that commute with all
the elements of $\cal A$. Clearly, ${\cal A}\subset {\cal A}''$ and ${\cal A}'=
{\cal A}'''$. If ${\cal A}= {\cal A}''$, we will say that ${\cal A}$ is a {\em
von Neumann's algebra}.
\end{definition}
The construction of an algebra by means of finite numbers of algebraic
operations on a basic set of bounded operators and its subsequent topological
closure by suitable limiting procedures requires a certain care in order to
avoid ambiguities that may rise from the possible non--commutativity of
limiting procedures. Among various possible kinds of topological closure, the
most convenient and interesting is the {\em closure in the weak topology}. It
consists, in practice, in taking the limit of a sequence of operators by
evaluating the limits of its matrix elements. In the following, we will assume
that the algebras generated by a set of bounded operators are closed in the
weak topology. The well--known theorem called of the {\em bicommutant} then
holds \cite{NEU0}.
\begin{theorem}
\label{bicomm}  If ${\cal A}$ contains the unit operator $I$ and is closed in
the weak topology then ${\cal A}'' = {\cal A}$.
\end{theorem}
Let us state without proof the following proposition.
\begin{theorem}
\label{UPgenera} ${\cal A} = {\cal A}''$ can be generated from all of its
projectors as well as from all of its unitary operators.
\end{theorem}
Therefore, ${\cal A}'$ is determined by the condition of commuting with all the
projectors of ${\cal A}$ or with all the unitary operators of ${\cal A}$.

Let us now study how the weak--closure property permits the conversion of the
partial ordering of the subalgebras of ${\cal I}$ in a complete lattice of
subalgebras.

\begin{definition}
Let ${\cal A}\subset {\cal I}$ and ${\cal B}\subset {\cal I}$ two von Neumann's
algebras. Let us indicate by ${\cal A}\vee{\cal B}$ the smallest von Neumann's
algebra that contains ${\cal A}$ and ${\cal B}$ as subalgebras, and by ${\cal
A}\wedge{\cal B}$ the greatest von Neumann's algebra contained in ${\cal A}$
and ${\cal B}$.
\end{definition}
It can be easily proved that ${\cal A}\wedge{\cal B}= {\cal A}\cap{\cal B}$,
where $\cap$ is the set--theoretic intersection.

Since both ${\cal A}$ and ${\cal B}$ contain the set $\{\alpha I\}$ formed by
the multiples of the unit operator $I$, we have $$\{\alpha I\}\subset {\cal
A}\wedge{\cal B}\subset {\cal A}\vee{\cal B}\subset {\cal I}\,.$$ Note that
$\{\alpha I\}'= {\cal I}$ e ${\cal I}{\,}'=\{\alpha I\}$, and that in the
partially ordered set of subalgebras the apex exchanges the adjective {\em
containing} with the adjective {\em contained} and {\em greater} with {\em
smaller}; in particular, it exchanges ${\cal A}\subset {\cal B}$ with ${\cal
B}'\subset {\cal A}'$. We arrive thereby to establish the following theorem
\begin{theorem}
\label{dualita} If  $I\in {\cal A}={\cal A}''$ and $I\in {\cal B}= {\cal B}''$,
then the following lattice--theoretic--ordering relations hold
\begin{eqnarray}
& & \{\alpha I\}\subset {\cal A}\wedge{\cal B}\subset {\cal A}\vee{\cal
B}\subset {\cal I}\,;\quad \{\alpha I\}\subset ({\cal A}\vee{\cal B})'\subset
({\cal A}\wedge{\cal B})'\subset {\cal I}\,;\nonumber
\end{eqnarray}
and the duality relations $({\cal A}\wedge{\cal B})'= {\cal A}'\vee{\cal
B}'\,,\quad({\cal A}\vee{\cal B})'= {\cal A}'\wedge{\cal B}'$.
\end{theorem}
In short, the apex works as an operator of {\em relative orthocomplementation}.

Let us now introduce the notions of {\em factor}  and {\em factorization}.
\begin{definition}
An algebra $\cal A$ of bounded operators containing $I$, but not necessarily a
von Neumann's algebra, is called a {\em factor} if ${\cal A}\wedge {\cal A}'=
\{\alpha I\}$ and $\cal A\vee {\cal A}'= {\cal I}$. If ${\cal A}$ is a von
Neumann's algebra, the condition ${\cal A}\wedge {\cal A}'= \{\alpha I\}$ is
sufficient, as the second equality can be derived from the first by duality.
In this case $\cal A$ and ${\cal A}'$ are called  {\em coupled factors}.
\end{definition}
A trivial example of coupled factors is provided by ${\cal A}={\cal I}$ and
${\cal A}'=\{\alpha I\}$. An elementary example is that formed by the direct
products of matrices $A\otimes I'_{N'}$ and $I_N\otimes A'$, where $A$ and $A'$
are squared matrices and $I_N$, $I'_{N'}$ are the unit matrices, respectively
of dimensions $N$ and $N'$. $N$ and $N'$ can be finite or infinite. The most
interesting cases, however, are those in which the operators of each factor
cannot be represented as direct products of the form indicated above. The
existence and the properties of these factors are just the subject of our
investigation.

\subsection{Isometric projectors}
Here, we introduce briefly a few important notions on partially isometric
operators \cite{NEU1}. Let $\cal K$ be a subspace of $\cal H$. In the
following, the subspace orthogonal to $\cal K$ in $\cal H$ we will be denoted
by ${\cal H}-{\cal K}$.
\begin{definition}
\label{isometria} An operator $\hat U$ is called {\em partially isometric} if
it maps a subspace ${\cal K}_1 \subset \cal H$ onto a subspace ${\cal K}_2
\subset \cal H$ leaving vector--norms unchanged and annihilating the subspace
${\cal H}- {\cal K}_1$. That is, if $f \in {\cal K}_1$ implies $\hat U f \in
{\cal K}_2$, with $(\hat U f, \hat Uf) = (f,f)$, while $f \in {\cal H}- {\cal
K}_1$ implies $\hat U f = 0$.
\end{definition}
From the symmetry of the definition it follows that the adjoint operator $\hat
U^\dag$ maps isometrically ${\cal K}_2$ into ${\cal K}_1$ annihilating the
orthogonal subspace ${\cal H}- {\cal K}_2$. We can easy verify that $\hat
U^\dag \hat U$ and $\hat U \hat U^\dag$ project $\cal H$ respectively onto
${\cal K}_1$ and ${\cal K}_2$. This fact leads naturally to the following
definition.
\begin{definition}
\label{isometrici} Two projectors $P_1$, $P_2$ will be called {\em isometric},
and we will write $P_1\sim P_2$, if there exists a partially isometric operator
$\hat U$ such that $\hat U^\dag \hat U = P_1$, $\hat U \hat U^\dag = P_2$. The
following equalities can be easily verified $\hat U P_1 \hat U^\dag = P_2$,
$\hat U^\dag P_2 \hat U = P_1$.
\end{definition}
In the next, we will use the following definition and soon after a theorem
based on the considerations reported along with the definition.
\begin{definition}
Two partially isometric operators $\hat U^{(1)}$ and $\hat U^{(2)}$ are called
orthogonal if $\hat U^{(1)\dag}\hat U^{(2)} = \hat U^{(2)} \hat U^{(1)\dag}
=0$.
\end{definition}
Define the projectors $$P^{(1)}_1=\hat U^{(1)\dag} \hat U^{(1)}\,,\quad
P^{(1)}_2= \hat U^{(1)}\hat U^{(1)\dag}\,,\quad P^{(2)}_1= \hat U^{(2)\dag}
\hat U^{(2)}\,,\quad P^{(2)}_2=\hat U^{(2)}\hat U^{(2)\dag}\,.$$ If $\hat
U^{(1)}$ and  $\hat U^{(2)}$ are orthogonal, then the equations
$$(\hat U^{(1)}+\hat U^{(2)})^\dag (\hat U^{(1)}+\hat
U^{(2)}) = P^{(1)}_1 + P^{(2)}_1\,,\quad (\hat U^{(1)}+\hat U^{(2)})(\hat
U^{(1)}+\hat U^{(2)})^\dag = P^{(1)}_2 + P^{(2)}_2$$ clearly hold. Moreover
$P^{(1)}_1 P^{(2)}_1=0$, $P^{(1)}_2 P^{(2)}_2=0$. In short, $\hat U^{(1)}+\hat
U^{(2)}$ is an isometric operator and $P^{(1)}_1 + P^{(2)}_1$, $P^{(1)}_2 +
P^{(2)}_2$ are projectors. Also the converse can be easily proved: if
$P^{(1)}_2 P^{(2)}_2=0$ and $P^{(1)}_1 P^{(2)}_1=0$, then $\hat U^{(1)}$ e
$\hat U^{(2)}$ are orthogonal. These results lead directly to the following
conclusion.
\begin{theorem}
\label{isorto} If for the projectors $P^{(1)}_1$, $P^{(1)}_2$, $P^{(2)}_1$,
$P^{(2)}_2$ the properties $$P^{(1)}_1P^{(1)}_2=0\,,\quad
P^{(2)}_1P^{(2)}_2=0\,,\quad P^{(1)}_1\sim P^{(1)}_2\,,\quad P^{(2)}_1\sim
P^{(2)}_2$$ hold, then $P^{(1)}_1+P^{(2)}_1$, $P^{(1)}_2+P^{(2)}_2$ are
projectors and $P^{(1)}_1+P^{(2)}_1\sim P^{(1)}_2+P^{(2)}_2$.
\end{theorem}

\subsection{Remarkable relationships between $\cal A$ and ${\cal A}'$}
Let us now evidence a few operator properties that are naturally related to the
notions introduced in the previous section. The theorems that we are on the
point to present belong to the repertoire of standard notions of the theory of
operators \cite{BRATT}. The first two of them are well known, and we wish to
recall them only for their importance in the theory of factorization that we
want to describe.

\begin{theorem}[of the root operator]
If $B$ is a non--negative definite and bounded operator, then the operator
$B^{1/2}$ belongs to the algebra generated by $I$ e $B$.
\end{theorem}
We can indeed easily prove the convergence of the binomial series
$$
B^{1/2} = [I+(B-I)]^{1/2}=\sum_{n=0}^\infty
\frac{\frac{1}{2}(\frac{1}{2}-1)\cdots (\frac{1}{2}-
 n+1)}{n!}(B-I)^n\,.
$$
\begin{theorem}[of the polar decomposition]
Every bounded  operator $A$ can be written as $A=\hat U|A|$, where $|A| =
(A^\dag A)^{1/2}$ is non--negative definite and $\hat U$ is partially
isometric.
\end{theorem}
Consequently, also the projectors  $P_A= \hat U^\dag \hat U$ and $P_{A^\dag}=
\hat U \hat U^\dag$ are determined. Let ${\cal D}(A)$ and ${\cal C}(A)$ be
respectively the domain and the range of $A$. Denote by $[{\cal D}(A)]$ the
closure of ${\cal D}(A)$, that is, the smallest subspace of $\cal H$ that
contains ${\cal D}(A)$, and with $[{\cal C}(A)]$ the closure of ${\cal C}(A)$.
Then $[{\cal D}(A)]= P_A {\cal H}$ and $[{\cal C}(A)]= P_{A^\dag} {\cal H}$.
\begin{theorem}[of the commuting projector]
\label{codominio} Let $A\in \cal A$ and $\hat P$ a projector of ${\cal I}$. A
necessary and sufficient condition that the equality $[A, P]=0$ holds is
${\cal C}(A\tilde P)\subset {\cal C}(\tilde P)$ and ${\cal C}(A^\dag\tilde
P)\subset {\cal C}(\tilde P)$.
\end{theorem}
The necessity of the condition is evident. As for the sufficiency, it is clear
that the inclusive relations  between the ranges are equivalent to $\hat P A
\hat P = A\tilde P$ and $\tilde P A^\dag \tilde P = A^\dag \tilde P$. The
comparison of the first equation with the adjoint of the second yields $[\tilde
P, A]=0$.

\begin{theorem}[of the crossed projector]
\label{incrociato} Let $P$ be a projector of $\cal A$ and $P'$ a projector of $
{\cal A}'$. Then $PP' = 0$ implies $P=0$ or $P'=0$. In other words, if both
$P$ and $P'$ differ from zero then $PP'\neq 0$.
\end{theorem}
Proof: Consider the subspace $\tilde {\cal K} $ formed by all vectors $\tilde
f \in \cal H$ such that $PX\tilde f =0$ for all the operators $X\in {\cal A}$.
In symbols, $ \tilde {\cal K} = \{\tilde f: PX\tilde f =0, X\in {\cal A}\}$.
Let $\tilde P$ be the projector defined by $\tilde P {\cal H}= \tilde {\cal
K}$. If $A \in {\cal A}$, the relation $A\tilde f\in \tilde {\cal K}$ holds as
$XA \in {\cal A}$; if $A' \in {\cal A}'$, $A'\tilde f\in \tilde {\cal K}$
holds as $PXA'\tilde f =A'PX\tilde f = 0$. In both cases, we have the
inclusive relations
$${\cal C}(A\tilde P)\subset {\cal C}(\tilde P)\,,\quad {\cal C}(A^\dag\tilde
P)\subset {\cal C}(\tilde P)\,,\quad {\cal C}(A'\tilde P)\subset {\cal
C}(\tilde P)\,,\quad {\cal C}(A'^\dag\tilde P)\subset {\cal C}(\tilde P)\,,$$
which enable the application of theorem \ref{codominio}. We have then $[\tilde
P, A] = [\tilde P, A']=0$. Since $\cal A$ e ${\cal A}'$ have in common only
multiples of the unit operator we obtain $\tilde P = \alpha I$, with $\alpha =
1$ or $\alpha = 0$, in other terms, $\tilde {\cal K}={\cal H}$ or $\tilde {\cal
K}=\varnothing$. Assume now $PP'=0$. This means that for every $f \in {\cal H}$
we have $P' f \in \tilde {\cal K}$ and $P f \in {\cal H} - \tilde {\cal K}$.
Since $\tilde {\cal K}={\cal H}$ or $\tilde {\cal K}=\varnothing$, we deduce
that at least one of the two projectors is zero.

\begin{theorem}[of local comparability]
\label{locale} Two non--zero projectors $P_1$ and $P_2$ of $\cal A$ possess
isometric segments. That is, there exist a projector $\Delta P_1 \subset P_1$
and a projector $\Delta P_2 \subset P_2$ such that $\Delta P_1 \sim \Delta
P_2$.\footnote{$P_1 \subset P_2$ means $P_1 P_2=P_1$.}
\end{theorem}
It is sufficient to prove that there exists a non--zero partially isometric
projector $\hat U_{12}$ such that $\hat U_{12}^\dag \hat U_{12}\subset P_1$ and
$\hat U_{12} \hat U_{12}^\dag\subset P_2$.

Assume $f\in P_1  {\cal H} \equiv {\cal K}_1$ and normalize $f$ so that
$(f,f)=1$. Form the space ${\cal K}'_f$ that comprises all vectors that can be
obtained  by applying all the operators of $\cal A$ to $f$, in symbols ${\cal
K}'_f = [{\cal A} f]$, and denote by $P'_f$ the projector that sends $\cal H$
into ${\cal K}'_f$. Clearly, if $A\in \cal A$ then ${\cal C}(AP'_f)\subset
{\cal C}(P'_f)$ and moreover ${\cal C}(A^\dag P'_f)\subset {\cal C}(P'_f)$,
since also $A^\dag\in \cal A$. Then, from Theor. \ref{codominio} we obtain
$[P'_f, A]=0$ for every $A \in {\cal A}$, and therefore $P'_f \in {\cal A}'$.
Consequently, we obtain not only $f \in P_1P'_f\neq 0$ by construction, but
also $P_2P'_f\neq 0$ by theorem \ref{incrociato}.

Let now be $g\in (P_2P'_f) {\cal H} = (P'_f P_2){\cal H}$ with $(g,g)=1$. This
means that for at least one operator $\hat A\in {\cal A}$ the inequality
$$\parallel P_2 \hat A f - g
\parallel\, \equiv\, \parallel  P_2  \hat A P_1 f -g\parallel <1$$ holds,
which implies $ P_2  \hat A P_1 \neq 0$. Pose $A_{12}= P_2 \hat A P_1$ and
form the  polar decomposition $A_{12}= \hat U_{12}|A_{12}|$. It can then be
easily proved that $\hat U_{12}$ has the expected properties.

\subsection{The first theorem of comparability}
Basing on Theor.\ref{locale}, we can carry out the following construction.
Assume $P_1, P_2 \in \cal A$, $\Delta P_1 \subset P_1$, $\Delta P_2 \subset
P_2$ and $\Delta P_1 \sim \Delta P_2$.  Define $\Delta^{(1)} P_1 \equiv \Delta
P_1 $ and $\Delta^{(1)} P_2 \equiv \Delta P_2$, with the purpose of starting
the enumeration of subsequent applications of the theorem. By subtracting the
isometric segments $\Delta P_1$ and $\Delta P_2$ respectively from $P_1, P_2$
we obtain two new projectors $P_1-\Delta P_1$, $P_2 -\Delta P_2$. If none of
these is zero we can apply again theorem \ref{locale}, thus finding two new
isometric segments $\Delta' P_1\sim \Delta' P_2$ respectively orthogonal to
the previous ones. Pose $\Delta^{(2)} P_1 = \Delta^{(1)} P_1 + \Delta' P_1$,
$\Delta^{(2)} P_2 = \Delta^{(1)} P_2 + \Delta' P_2$. Then, from theorem
\ref{isorto}, we obtain $\Delta^{(2)} P_1 \sim \Delta^{(2)} P_2$. We can
iterate this procedure possibly until one of the two remainders becomes zero.
Alternatively, the procedure continues indefinitely. Even in this case, we can
exhaust one of the remainders by invoking the principle of Transfinite
Induction. Pose $ \bar P_1 = \sum_n \Delta^{(n)} P_1$, $\bar P_2 = \sum_n
\Delta^{(n)} P_2$. Since all segments of each projector are mutually
orthogonal, we have $\bar P_1 \sim \bar P_2$. We arrive in this way to state
the following  {\em comparability} theorem \cite{NEU0}.
\begin{theorem}
If $P_1, P_2$ are two projectors of a same factor $\cal A$, then either there
exists a $\bar P_1$ such that $P_1\sim \bar P_1 \subset P_2$ or a $\bar P_2$
such that $P_2\sim \bar P_2 \subset P_1$. In the first case we will write
$P_1\preceq P_2$, in the second $P_1\succeq P_2$.
\end{theorem}
Clearly, $P_1\preceq P_2$ e $P_2\preceq P_1$ imply $P_1\sim P_2$.

In conclusion, the theorem asserts the possibility of mapping one into the
other all the projectors of a same factor, which implies an unexpected {\em
homogeneity} of the algebraic structure of the factor, in particular, of the
spectral densities of the projectors. The reason of this homogeneity lies in
that the factorization determines the splitting of the algebra $\cal I$ in two
subalgebras, which, although independent of each other, remain mutually
correlated by a common dimensional constraint. Note indeed that to prove this
comparability theorem we had to involve the projectors of the commuting
algebra. This correlation is for certain aspects similar to that determined by
a anholonomic constraint between two parts of a classical system. As the state
of the system changes, the two parts roll and crawl onto each other in such a
way that each of them can reach any desired configuration independently of the
other, leaving however unchanged the dimension of the contact.

In certain aspects, this theorem is the analog of Cantor--Bernstein's theorem
on comparability of sets, which stands at the basis, firstly, of the notion of
{\em cardinality}, secondly, of that of {\em measurability}. If we consider
that the projectors of quantum--mechanical systems are the analog of the sets
of states of classical systems, the question arises quite naturally of what may
be, in the world of projectors, the analog of the measure of a set.

The existence of a relation of total ordering in the isometry class of
projectors, together with the additive property of orthogonal projectors,
suggests the introduction of an additive measure $D(P)$, to be called the {\em
relative dimension} over the set of the projectors $P$ of a same factor. To be
consistent with the measure analog we impose the following conditions: (i)
$D(0)=0$; (ii) if $P_1P_2 =0$ then $D(P_1+P_2)=D(P_2)+D(P_1)$; (iii)
$P_1\preceq P_2$ is equivalent to $D(P_1)\le D(P_2)$.

Using the perseverance of additive property across isometric maps, carried out
in all possible (finitary or infinitary) ways, among the projectors $P\in \cal
A$, we arrive to the following important result:
\begin{theorem} The relative dimensions are exhausted by the following three types of
possibilities:
\begin{itemize}
\item[\bf I.] $D(P)= 0, 1, \dots, n$ {\em (type $I_n$)}; or $D(P)= 0, 1,
\dots, \infty$ {\em (type  $I_\infty$)}.
\item[\bf II.] $D(P)$ takes a value in the interval $[0, c]$,
where $c$ is a positive real number {\em (type $II_c$)}; or in the interval
$[0, \infty]$ {\em (type $II_\infty$)}.
\item[\bf III.]  $D(P) = 0, \infty$ {\em (type $III$)}.
\end{itemize}
\end{theorem}
Since in case $II_c$ the dimensions are defined up to an arbitrary scale
factor, we can pose $c=1$ and denote type $II_c$ as $II_1$.

As an example of type $I_\infty$, we can indicate the algebra $\cal I$ itself.
In this case, all projectors of finite dimension have a discrete spectrum. This
means that the projectors whose spectrum is continuous or partially continuous
have infinite dimension. By contrast, in factors $II_1$ and $II_\infty$, the
sole projector with a discrete spectrum is zero, the spectra of all others
being continuous. This is just the reason why the dimensions of these types of
projectors have finite ratios.

\subsection{The second theorem of comparability}
We will now establish a one--to--one mapping between a class of projectors of
${\cal A}$ and a class of projectors of ${\cal A}'$ by the following
construction.

Assume $f\in\cal H$. Denote by $[{\cal A}f]$ the smallest subspace of $\cal H$
that contains the vectors that can be obtained by the application of all the
operators $A\in {\cal A}$ to $f$, and by $[{\cal A}'f]$ that obtained by the
application of all $A' \in {\cal A}'$ to $f$. Denote by $P'_f$ the projector
that sends $\cal H$ onto $[{\cal A}f]$ and with $P_f$ that sends  $\cal H$
onto $[{\cal A}'f]$. We wish to prove that $P_f \in {\cal A}$ and $P'_f\in
{\cal A}'$.

In fact, all the unitary operators $U\in {\cal A}$ leave invariant $[{\cal
A}f]$ as, if $A\in {\cal A}$, also $UA\in {\cal A}$ and consequently
$UP'_fU^\dag = P'_f$. In the same way, all the unitary operators $U'\in {\cal
A}'$ leave invariant $[{\cal A}'f]$ and consequently $U'P_fU'^\dag = P_f$.
Since ${\cal A}$ and ${\cal A}'$ can be generated from their unitary operators
(theorem \ref{UPgenera}), the commutation relations $[P'_f, {\cal A}]=0$ and
$[P_f, {\cal A}']=0$ do hold. We can then say that every vector $f\in\cal H$
determines a pair of projectors: $P_f\in {\cal A}$ and $P'_f\in {\cal A}'$.

Let now $P_{f_1}$, $P_{f_2}$, $P'_{f_1}$, $P'_{f_2}$ be the projectors
determined by the vectors  $f_1, f_2\in \cal H$  as indicated above. Then,
because of the first comparability theorem, $P_{f_1}$ and $P_{f_2}$ are
comparable and such are also $P'_{f_1}$ and $P'_{f_2}$. It is not said,
however, that the first ones be compatible with the second ones. The following
theorem, which we limit ourselves to state because its proof is rather
cumbersome, however holds
\begin{theorem}
If $P_{f_1}\preceq P_{f_2}$, then $P'_{f_1}\preceq P'_{f_2}$. Moreover, for
every $P\in {\cal A}$ there exists a vector $f$ for which $P_f \sim P$ and for
every $P'\in {\cal A}'$ there exists a vector $f$ for which $P'_f \sim P'$.
\end{theorem}
Clearly, the first part of the theorem implies the correspondence
$$D(P_{f_1})\lesseqgtr D(P_{f_2})\rightleftarrows D'(P'_{f_1})\lesseqgtr
D'(P'_{f_2})\,,$$ where $D'(P')$ is the relative dimension of $P'\in {\cal
A}'$. This means that the range of $D(P_f)$, $f\in \cal H$, can be mapped onto
that of $D'(P'_f)$, $f\in \cal H$, and/or vice versa. But then, from the second
part of the theorem, also the range of $D(P)$, $P\in {\cal A}$, can be mapped
onto that of $D'(P')$, $P'\in {\cal A}'$, and/or vice versa. We deduce that
${\cal A}$ and ${\cal A}'$ belong to the same type and that the dimensions of
their projectors exhaust the following possibilities:
\begin{itemize}
\item[{\bf I.}] $D(P)=0, 1, 2, \dots , n \le \infty$; $D'(P')=0, 1, 2, \dots ,n' \le \infty$.
\item[{\bf II.}] $0 \le D(P) \le c\le \infty$; $0 \le D'(P') \le c'\le \infty$.
\item[{\bf III.}] $D(P), D'(P')=0, \infty$.
\end{itemize}
If $n=n'$ e $c=c'$, we will say that the factors are {\em symmetrically
coupled}.

%\newpage
\section{Discrete systems}
\label{assiomi} The class of systems that we will now describe provides some
examples of {\em  symmetrically coupled factors} for all types indicated in the
previous section. The class is a rather wide. It is based on the construction
of a Hilbert space of measurable functions in which a maximal algebra of
projectors is automorphically transformed by a {\em discrete} group of unitary
operators. Because of this, the systems will be defined {\em discrete}. More
general classes of systems could be obtained by introducing finite or infinite
continuous groups. The reason of the restriction to discrete systems, however,
will be soon apparent.

\subsection{Functional Hilbert spaces of maximum generality}
Let ${\cal H}_X$  be the Hilbert space of measurable functions $f(x)$, with $x$
running over a set of points $X$. To extend the notions of {\em measurable set}
and {\em measurable function} to the case in which $X$ is not a topological
set, we introduce the following notion of {\em measure}.

For every subset $S\subset X$ we define a real function $\mu(S)$ characterized
by the following axioms:
\begin{itemize}
\item[(i)] {\em Positivity}: $0 \le \mu(S)\le \infty$.
\item[(ii)] {\em Ordering}: $S \subset S'$ implies $\mu(S) \le \mu(S')$.
\item[(iii)] {\em Additivity}: For every finite or infinite collection $S_1, S_2,
\dots$,
$$\mu(S_1\cup S_2\cup \dots) \le \mu(S_1)+ \mu(S_2)+ \dots\,.$$
\end{itemize}
We adopt now the following definition\footnote{In the following, the
complementary set of $S$ in $X$ will be denoted by $\bar S$ and the symmetric
difference $(S\cup S')\cap (\bar S \cup \bar S')$ with $S-S'$ (then $\bar S
\equiv X-S$).}.
\begin{definition} {\em Misurability according to Carath\éodory:}
\item The subset $S$ is called {\em measurable} if and only if for every subset
$S'\subset X$ the equality
$$\mu(S') =  \mu(S' - S\cap S')+\mu(S\cap S')$$
holds.
\end{definition}
Let us now continue the list of axioms:
\begin{itemize}
\item[(iv)] {\em Covering}: If $\mu(S_0)< \alpha$ then there exists a measurable set $S\supset
S_0$ con $\mu(S)< \alpha$.

\item[(v)] {\em Separability}: There exists a countable collection of
measurable subsets $S^{(1)}, S^{(2)}, \dots$ of finite measure and $\bigcup_i
S^{(i)}=X$, such that if $x, y \in S^{(i)}$ or $x, y \notin S^{(i)}$ for all
$i$, then $x=y$.\footnote{The original formulation reads: If $x\in S^{(i)}$ is
equivalent to  $y\in S^{(i)}$ for all $i$ then $x=y$. The condition $\bigcup_i
S^{(i)}=X$ is superfluous. Indeed, if it is not satisfied and the measure of
$S_0= X-\bigcup_i S^{(i)}$ is finite, we can add $S_0$ to the collection so as
to have $\cup_i S^{(i)}=X$. If on the contrary $\mu(S_0)= \infty$ then $S_0 =
\{x_0\}$ (one-point set). But in this case every measurable function $f(x)$
must satisfy the condition $f(x_0)=0$. Therefore, the point  $x_0$ can be
omitted from $X$.}
\end{itemize}

The latter axiom serves to replace the (absent or ignored) notion of {\em
topological separability}. In case $X$ is an Euclidean space of finite
dimension, we can take as $S^{(i)}$ all the spheres of rational coordinates and
radii.

\begin{definition} {\em Generalized notion of  Lebesgue measurability:}
\item A complex function $f(x)$ defined at all points $x \in X$ is called
{\em measurable} if for every real number $\alpha$ the sets
$$S_\alpha =\{x; \Re[f(x)]>\alpha\}\,,\quad S'_\alpha =\{x;
\Im[f(x)]>\alpha\}$$ are measurable.
\end{definition}
On this basis, we can introduce a generalized notion of Lebesgue integral.
Accordingly, we will define in ${\cal H}_X$ the scalar products
$$ (f_1,  f_2) = \int_X f^*_1(x)f_2(x)\,d\mu(x)\,.
$$
It is understood that two functions $f(x)$ and $f'(x)$ with square--integrable
moduli differing only over a zero--measure set represent the same vector of
${\cal H}_X$. In this case we will write $f(x)=f'(x)$ {\em a.e.} ({\em almost
everywhere}).

An important class of measurable functions is that of {\em box functions}:
$\chi_S(x) = 1$ for $x \in S$  and $\chi_S(x) = 0$ for $x \notin S$, where $S$
is a measurable subset of $X$. We will then have
$$
\int_X \chi_S(x)\,d\mu(x)= \int_S \,d\mu(x)=\mu(S) \,.
$$
It is known from  Lebesgues' theory of integration that box--function set is
dense in the space of measurable functions. This means that every measurable
$f(x)$ is {\em a.e.} equal to a limit of linear combinations of box functions.
Shortly, we will say that every $f(x)$ can be {\em a.e. generated} by box
functions.

\subsection{Discrete groups that are free and ergodic}
Let $G$ be a finite or infinite discrete group, whose elements $g$ operate on
$X$ by point transformations $x'=gx$. These transform a subset $S$ of $X$ to a
subset $S'= gS \equiv\{gx; x\in S\}$.

Let add now the following axioms (Murray and von Neumann, 1936; von Neumann,
1940).
\begin{itemize}
\item[(vi)] {\em Closure of measurability with respect to $G$:} If $S$ is measurable, also $gS$ is measurable.
\item[(vii)] {\em Freedom:} For every $g\in G$, the set $S_g=\{x: gx = x\}$
 has measure zero. This notion generalizes the condition that there are no fixed points in case $X$ is discrete.
\item[(viii)] {\em Ergodicity:} From $g\neq 1$ and $\mu(gS-S)=0$, where
$S-S'$ denotes the symmetric difference $S\cup S'-S\cap S'$, it follows either
$\mu(S)=0$ or $\mu(S)=\mu(X)$.
\end{itemize}

The axiom just listed characterize the collection of measurable subsets of $X$
as a Borelian system that is {\em separating} and {\em closed} with respect to
$G$.

Finally, let us add, only temporarily however, an axiom whose elimination will
lead, as we will see, to interesting consequences.
\begin{itemize}
\item[(ix)] {\em Measure invariance:}  The measure $\mu$ is invariant with respect to $G$,
that is, $\mu(S)=\mu(gS)$.
\end{itemize}
Therefore, in this case, for every measurable $f(x)$ and every $g\in G$ we
have the equations
$$
\int_X f(x) \,d\mu(gx)=\int_X f(g^{-1}x)\,d\mu(x)= \int_X f(x)\,d\mu(x)\,.
$$

\subsection{The algebras of discrete systems}
\label{algdiscr} We pass now to describe the algebraic structure of the system
that implements in the most natural way the axioms listed above.
\begin{definition}
Define in ${\cal H}_X$ the operators $$L_{\phi(x)} f(x) = \phi(x) f(x)\,,$$
where $\phi(x)$ is a bounded and measurable complex function of $x$.
\end{definition}
Clearly, these operators are bounded and if $\phi(x)= \phi'(x)$ {\em a.e.},
$L_{\phi(x)}$ and $L_{\phi'(x)}$ define the same operator. They form a
commutative algebra that will be denoted by $\cal L$.

Note that the operators $P_S\equiv L_{\chi_S(x)}$, where $\chi_S(x)$ are box
functions, form a commutative family of projectors, which in the following
will be denoted by $\cal P$. These satisfy the relations $P_{S} + P_{S'}=
P_{S\cup S'} + P_{S\cap S'}$ for all pairs of measurable sets $S, S'$. In
particular, $P(S)+P(\bar S)= I$ (the {\em unit operator}).

Note moreover that the operators $V_{\lambda(x)} = L_{\exp i \lambda(x)}$,
where $\lambda(x)$ is a real bounded and measurable real function of $x$,
generate an Abelian group, which will be denoted by $\cal V$.

Note finally that, since every measurable function $f(x)$ can be a.e.
generated by box functions, both $\cal L$ and $\cal V$ can be generated by
$\cal P$.

\begin{definition}
A commutative algebra $\cal L$ is called {\em maximal} if it satisfies the
equation ${\cal L} = {\cal L}'$.
\end{definition}

\begin{definition}
A family of projectors $\cal P$ is called maximal if from $[P, P_S]=0$ for any
$P_{S}\in \cal P$ it follows $P\in \cal P$.
\end{definition}
Basing on these definitions we prove the following theorem:
\begin{theorem}
\label{mass} The family of projectors  $\cal P$ is maximal. Consequently,
${\cal L}$ is a maximal commutative algebra.
\end{theorem}
Proof: Firstly, note that, since $\cal L$ can be generated by $\cal P$, if $P$
commutes with every $P_{S}\in \cal P$, it commutes also with every
$L_\phi(x)\in\cal L$. We have therefore $P L_{\phi(x)}f(x) = L_{\phi(x)}P
f(x)$ and consequently
\begin{equation}
\label{Pphi} P[\phi(x)f(x)] = \phi(x)[Pf(x)]\,.
\end{equation}
Secondly, note that, since $P$ is a bounded operator, the functions $P f(x)$
are measurable. Now, from the collection $S^{(1)}, S^{(2)}, \dots$ introduced
with axiom (v) we can form the collection
$$T^{(i)} = S^{(i)} - \bigcup_{j=1}^{i-1}S^{(j)}\,.$$
It can be easily proved that $\mu(T^{(i)}) < \infty$, $T^{(i)}\cap T^{(j)} =
\varnothing$ (the empty set) and $ \cup_i T^{(i)} = X$.

Let us then define the pairwise box--functions $e_i(x) =1$ if $x\in T^{(i)}$
and $e_i(x) =0$ if $x\in \bar T^{(i)}$. It is evident that $e_i(x)^2 = e_i(x)$,
$\sum_i e_i(x)=1$ and that the projectors $P_i\equiv P_{T^{(i)}}$, clearly
belonging to $\cal P$, operate on the functions $f(x) \in {\cal H}_X$ as
follows:
$$P_i f(x) = e_i(x)f(x)\,.$$
Therefore, from $[P P_i - PP_i] \,e_i(x) = 0$ we obtain  $P e_i(x) = e_i(x)P
e_i(x)$. Consequently, since for $i\neq j$ we have $e_i(x)\, e_j(x)=0$, there
is no loss of generality in posing $P \,e_i(x) = \phi(x)\, e_i(x)$ for all $i$.
Moreover, since $P^2 = P$, the equation $\phi(x) = \phi(x)^2$ holds. This means
that the possible values of $\phi(x)$ are 0 and 1.

Now let $S'$ be the set on which $\phi(x)=1$ and write $\phi(x)= \chi_{S'}(x)$.
Then, we can also write $P \,e_i(x) = \chi_{S'}(x)\, e_i(x)$. Using
(\ref{Pphi}) we obtain
$$P [\phi(x)e_i(x)] =\phi(x) [Pe_i(x)] = \phi(x)\chi_{S'}(x)\,e_i(x)\,,$$
from which, summing over $i$ and posing $\phi(x) = f(x) \in {\cal H}_X$, we
obtain $P f(x) = \chi_{S'}(x)f(x)$. Since $Pf(x)$ and $f(x)$ are measurable and
$\chi_{S'}(x)$ is bounded, also this function is measurable. Therefore $S'$ is
measurable and then $P\equiv P_{S'} \in \cal P$.
\begin{definition}
Now define the operators $$U_g f(x) = f(gx)\,.$$
\end{definition}
Note that their product follows the rule $U_g U_{g'} = U_{g'g}$. These form a
group of unitary  operators that will be indicated by $\cal U$. Indeed, from
$d\mu(g^{-1}x)= d\mu(x)$ we obtain
\begin{eqnarray}
(U_g f_1, U_g f_2)   =   \int_X f^*_1(gx)f_2(gx)\,d\mu(x)=\int_X
f^*_1(x)f_2(x)\,d\mu(g^{-1}x) = (f_1,  f_2) \,. \nonumber
\end{eqnarray}
Also note that
\begin{eqnarray}
U_g L_{\phi(x)} U_g^{-1} f(x) = U_g L_{\phi(x)} f(g^{-1}x)= U_g \phi(x)
f(g^{-1}x)= \phi(gx) f(x)= L_{\phi(gx)} f(x)\,.\nonumber
\end{eqnarray}
Consequently, because of the arbitrariness and completeness of $f(x)$, we can
also write
$$
 U_g L_{\phi(x)}U_g^{-1} = L_{\phi(gx)}\,.
$$
In other terms, the unitary operators $U_g$ induce a group of automorphisms on
$\cal L$ and then also on $\cal P$ e $\cal V$. In particular, we have
\begin{equation}
\label{UPU}
 U_g P_S U_g^{-1} = P_{gS}\,.
\end{equation}

Let us now prove two important theorems that will be used in the following
section.

\begin{theorem}
\label{free} The equation $L_{\phi(x)}= L_{\psi(x)} U_g$ is possible only if
either $g= 1$ and $\phi(x)=\psi(x) \neq 0$ hold a.e., or if $g\neq 1$ e
$\phi(x), \psi(x) = 0$ a.e.
\end{theorem}
Proof: This theorem is a consequence of group--freedom axiom (vii). To prove
this we start from noting that the hypothesized equation is equivalent to
$\phi(x)f(x) =\psi(x) f(gx)$ {\em a.e.} for every $f(x)\in {\cal H}_X$.

Using the sets $T^{(i)}$ and the measurable functions $e_i(x)$ introduced in
the proof of theorem \ref{mass}, we obtain the equations $\phi(x)\,e_i(x)
=\psi(x)\, e_i(gx)$ {\em a.e.}, then $\phi(x) =\psi(x)$ {\em a.e.} in
$T^{(i)}\cap g T^{(i)}$. Hence, summing over $i$ and noting that$\bigcup_i
T^{(i)}= \bigcup_i g T^{(i)}=X$, we deduce $\phi(x) =\psi(x)$ {\em a.e.}, $x\in
X$. Therefore the hypothesized equality translates to the equation
\begin{equation}
\label{intphi1} \int_X \vert \phi(x)\vert^2 \vert f(x)- f(gx)\vert^2 d\mu(x) =
0
\end{equation}
with $g\neq 1$.

Using the sets $S^{(i)}$ introduced in axiom (v), take the box functions $f(x)$
defined by $\hat e_i(x)= 1$ if $x\in  S^{(i)}$  and $\hat e_i(x)= 0$ if $x\in
\bar S^{(i)}$. Then in place of (\ref{intphi1}) we have
\begin{equation}
\label{intphi2} \int_X \vert \phi(x)\vert^2 [\hat e_i(x)-\hat e_i(gx)]d\mu(x) =
\int_{S^{(i)}-g^{-1}S^{(i)}} \vert \phi(x)\vert^2 d\mu(x) = 0\,. \nonumber
\end{equation}
Once posed $X_0 = \cup_i (S^{(i)}-g^{-1}S^{(i)})$, these imply
$$
\int_{X_0} \vert \phi(x)\vert^2 d\mu(x) \,\le \,\sum_i
\int_{S^{(i)}-g^{-1}S^{(i)}} \vert \phi(x)\vert^2 d\mu(x)= 0\,.
$$
This means that $\phi(x)$ does not vanish {\em a.e.} only in $\bar X_0= \cap_i
(\bar S^{(i)}-g^{-1}\bar S^{(i)})$, that is either for both $x$ and $g^{-1}x$
belonging to $\bar S^{(i)}$ (and to $g^{-1}\bar S^{(i)}$) or both belonging to
$S^{(i)}$ (and to $g^{-1} S^{(i)}$). But then, the separability axiom (v)
ensures that $x=gx$ for every $x\in X_0$ and $g\neq 1$. Moreover, the
group--freedom axiom (vii) ensures $\mu(X_0)=0$ and consequently $\phi(x)=0$
a.e.

Note that, by posing $\phi(x)=1$ in Eq.(\ref{intphi1}), the same line of
reasoning leads to the following result
\begin{corollary}
\label{fcost} The  equations $f(x)= f(gx)$ hold $a.e.$ for all elements $g\in
G$ if and only if $f(x)$ is a.e. constant.
\end{corollary}

The following theorem is a consequence of the ergodicity axiom (viii):
\begin{theorem}
\label{ergodic} Let ${\cal U}'$ be the commutant of ${\cal U}$. Then ${\cal
L}\cap {\cal U}' = \alpha I$ (multiples of the unit operator).
\end{theorem}
Proof: Since ${\cal L}$ can be generated by ${\cal P}$, it suffices to prove
that ${\cal P}\cap {\cal U}' = I$. Clearly, $I$ belongs to both ${\cal U}'$ and
${\cal P}$; in the first case because $I$ commutes with every $U_g$, in the
second case because $I =P(X)$. Assume absurdly that ${\cal P}\cap {\cal U}'$
contains a $P_S\neq 0, I$. Consequently $ U_g P_S U_g^{-1} = P_{gS}$ and, since
$P_S \in {\cal U}'$, the equality $P_S=P_{gS}$ must hold for every $g\in G$.
This means $S = gS$ {\em a.e.}, then $\mu(S - gS) = 0$ for every $g$. From the
ergodicity axiom (viii), this is possible only if either $\mu(S)=0$ or
$\mu(X-S)=0$, that is, if either $P_S=P_\varnothing =0$ or $P_S = P_X=I$. But
this means that
\begin{corollary}
The representation of the algebra generated by the operators $P_S$ and $U_g$ is
irreducible.
\end{corollary}

\section{The space of operational states}
In the previous section, it was built in a Hilbert space ${\cal H}_X$ a maximal
family $\cal P$ of commuting projectors $P_S\in \cal P$ indexed by the
measurable subsets $S$ of a set $X$. This family can be physically interpreted
as a complete set of compatible observables. This projector algebra is
equipped with a group of automorphisms $g\in G$ that operate freely and
ergodically on the subsets of $X$. The unitary operators $U_g$ that implement
the automorphism $U_g P_S U_g^{-1} = P_{gS}$ can be interpreted as the
representations of the operations $g$ that an external observer can perform on
the states of the system.

If these operations are carried out in the macroscopic world, we must assume
that they behave as a set of classical possibilities, i.e., like a set of
mutually exclusive results of a measurement. From the quantum mechanical point
of view, this means that they must be represented by a set of pairwise
orthogonal vectors $\vert g_1 \rangle, \vert g_2 \rangle,\dots$ of a Hilbert
space ${\cal H}_G$ indexed by the elements $g_i \in G$. We can normalize these
states so as to have $\langle g_i\vert g_j\rangle = \delta_{ij}$.

The operator algebra of this space exhibits some interesting aspects. The
inversion operator $Q$, defined by the equation $Q\vert g \rangle= \vert
g^{-1} \rangle$, implies the existence of two groups of operators: ${\cal
G}_L$ ({\em the left group}) and ${\cal G}_R$ ({\em the right group}), whose
elements $L_g\in {\cal G}_L$ and $R_g\in {\cal G}_R$ operate on the states of
${\cal H}_G$ as follows:
$$L_g\vert g_i\rangle = \vert gg_i \rangle\,,\,\, R_g\vert g_i\rangle = \vert g_ig
\rangle\,.$$ Indeed, from
$$QL_g\vert g_i \rangle = Q \vert gg_i \rangle=  \vert g_i^{-1} g^{-1}\rangle
=  R_{g^{-1}} \vert g_i^{-1}\rangle =R_{g^{-1}} Q\vert g_i\rangle$$ and from
$Q^2=1$ we obtain $L_g= QR_{g^{-1}}Q$. It can be easily proved that $[{\cal
G}_L, {\cal G}_R]=0$ and that the operators $R_g, L_g$ e $Q$ are unitary.

It is useful to define also the following operators:
$$ P_g =\vert g\rangle \langle g\vert\,,\,\, Q_g = \vert g^{-1}\rangle\langle g\vert\,\,.$$
Clearly, $P_g$ is a projector and $Q_g$ can be defined an {\em inversor}. The
following equations can be easily proved
$$Q^\dag_g = Q_{g^{-1}}\,,\quad Q_g Q_{g'} = P_{g^{-1}}\delta_{g^{-1} g'}\,,\quad Q = \sum_i Q_{g_i}\,.$$

\section{Hybrid spaces}
% SISTEMI DIALETTICI
The arguments introduced in the previous section ultimately suggest that the
quantum--mechanical representation of an observable system that can be
automorphically transformed by an external agent, combined with the
representation of the actions that can be performed by the agent, takes the
form of the direct product
$$
{\cal H}= {\cal H}_X\otimes {\cal H}_G\,,
$$
which in the following will be called a {\em hybrid space}. The vectors $\vert
\Phi \rangle \in {\cal H}$ can then be physically interpreted as the states of
the observed--observer system. They can be written as
$$
\vert \Phi \rangle = \hbox{$\sum_i$} f(g_i, x)\otimes \vert g_i \rangle\,,
$$
so that their scalar products are defined by the relationships
$$
\langle \Phi_1\vert \Phi_2 \rangle = \hbox{$\sum_i$} \int_X f_1^*(g_i, x)
f_2(g_i, x)\,d\mu(x)\,.
$$

\noindent {\bf Note}: {\em The elements $g_i \in G$, on which the functions
$f(g_i, x)$ appear to depend, could be safely replaced by the indices $i$. We
prefer, however, to exhibit the dependence on $g_i$ as this will allow us to
exploit identities of the sort}
$$
\hbox{$\sum_{i}$} a_{g_i}  b_{g_i} \equiv \hbox{$\sum_{i}$} a_{gg_i} b_{gg_i}
\equiv \hbox{$\sum_{i}$} a_{g_ig} b_{g_ig} \equiv \hbox{$\sum_{i}$}
a_{g_i^{-1}} b_{g_i^{-1}} \,,\,\, \hbox{etc.}
$$

Let us now introduce a different but particularly useful way to represent the
states of $\cal H$. Once the basis of ${\cal H}_G$ is chosen, any generic
vector $\vert \Phi \rangle\in {\cal H}$ is one--to--one to the ordered
collection $ f(g_1, x), f(g_2, x), \dots$, $g_i \in G$, of the vectors of
${\cal H}_X$. We can therefore indicate this equivalence by writing
$$\vert \Phi \rangle \sim \dvert f(g_1, x), f(g_2, x), \dots\drangle\equiv \dvert f(g_i, x)\,\drangle\,.$$
The following theorem indicates how this sort of representation can be
extended to the operators of ${\cal H}$.
\begin{theorem}
\label{Aij} The basis $\vert g_i \rangle$ being fixed, every bounded operator
$A$ of ${\cal H}$ is one--to--one with a matrix  $A_{g_i}^{g_j}$ of operators
of ${\cal H}_X$ satisfying the equations
$$
A \vert \Phi \rangle = \vert \Phi' \rangle \sim \dvert f'(g_i, x)\,\drangle
=\dvert\hbox{$\sum_{j}$} A_{g_i}^{g_j} f(g_j, x)\,\drangle\,.
$$
\end{theorem}
In other terms, the operator $A$ is one--to--one with the doubly ordered
collection of the operators $A_{g_i}^{g_j}$ of ${\cal H}$. We can therefore
indicate this fact by writing
$$ A \sim \dvert A_{g_i}^{g_j}\,\drangle$$
and abridge the action of $A$ on  $\vert \Phi \rangle$ by the equation
$$
A f(g_i, x) = f'(g_i, x) = \hbox{$\sum_{j}$} A_{g_i}^{g_j}f(g_j, x)\,,\quad
g_i, g_j\in G\,.
$$

The proof of this equation can be easily obtained by using the projectors $\bar
P_{g_i}$ of $\cal H$ defined by the equations $\bar P_{g_i}\vert \Phi\rangle =
f(g_i, x)\otimes \vert g_i \rangle$, that is the operators $\bar P_{g_i} \equiv
1\otimes P_{g_i}$, then by realizing that the operators $A_{g_i}^{g_j} $ of
${\cal H}_X$ defined by the equations
$$[ A_{g_i}^{g_j} f(g_j, x)]\otimes \vert
g_i \rangle = \bar P_{g_i}A \bar P_{g_j}\vert \Phi\rangle$$ satisfy the
required properties. Also the relationships
\begin{eqnarray}
\label{proggA}  \lambda A \sim \, \dvert \lambda
A_{g_i}^{g_j}\,\drangle\,;\quad A^\dag \sim \, \dvert
({A_{g_i}^{g_j}})^\dag\drangle\,; \quad A + B \sim \, \dvert A_{g_i}^{g_j}
+B_{g_i}^{g_j}\,\drangle\,;\quad AB \sim\, \dvert \hbox{$\sum_k$} A_{g_i}^{g_k}
B_{g_k}^{g_j} \,\drangle \,;
\end{eqnarray}
can be easily proved. In the following, $\dvert f(g_i, x)\drangle$ and $\dvert
A_{g_i}^{g_j} \drangle$ will be respectively called the projections of $\vert
\Phi \rangle$ and  $A$ in ${\cal H}_X$.

\subsection{The algebra of a hybrid space}
\begin{definition} Let us build in ${\cal H}={\cal H}_X\otimes {\cal H}_G$ the following
operators
\begin{equation}
\label{barLU}
 \bar L_{\phi(x)} = L_{\phi(x)} \otimes 1\,;\quad \bar U_g = U_g \otimes
L_{g^{-1}}\,.
\end{equation}
\end{definition}
They act on $\vert \Phi \rangle= \sum_i f(g_i, x)\otimes \vert g_i\rangle\
\equiv \sum_i f(gg_i, x) \otimes \vert gg_i\rangle$ in the following way
$$
\bar L_{\phi(x)} \vert \Phi\rangle = \sum_i \Bigl[L_{\phi(x)} \otimes 1\Bigr]
f(g_i, x) \otimes \vert g_i \rangle = \sum_i \phi(x) f(g_i, x) \otimes \vert
g_i\rangle \,;
$$
$$
\bar U_g \vert \Phi\rangle =  \sum_i \Bigl[U_g \otimes L_{g^{-1}}\Bigr] f(gg_i,
x) \otimes \vert gg_i\rangle= \sum_i f(gg_i, gx) \otimes \vert g_i\rangle\,.
$$
In the abridged form, we can write
$$
\bar L_{\phi(x)} f(g_i, x) =\phi(x) f(g_i, x)\,,\,\, \bar U_g f(g_i, x)  = f(g
g_i, g x)\,,\,\, g_i\in G\,.
$$
The operators $\bar L_{\phi(x)} $ are isomorphic to the operators $L_{\phi(x)}$
introduced in ${\cal H}_X$ and for them the already established theorems hold.
As for the operators $\bar U_g $, they are unitary because they are direct
products of unitary operators. Note that the automorphisms
\begin{equation}
\label{iso1} \bar U_g \bar L_{\phi(x)}\bar U_g^{-1} = \bar L_{\phi(gx)}\,,\quad
\bar U_g \bar P_S \bar U_g^{-1} = \bar P_{gS}
\end{equation}
have the same structure of the analogous operators defined in ${\cal H}_X$.

In the following, the commutative algebra formed by $\bar L_{\phi(x)}$ will be
indicated by $\bar {\cal L}$, the group of operators $\bar U_{g}$ will be
indicated by $\bar {\cal U}$ and the family of projectors $\bar P_S = \bar
L_{\chi_S(x)}$ will be  indicated by $\bar {\cal P}$.

\begin{definition}
Now define the operators
\begin{equation}
\label{barLUprime} \bar L'_{\phi(x)} = \hbox{$\sum_{i}$} L_{\phi(g^{-1}_i
x)}\otimes P_{g_i}\,;\quad \bar U'_g = 1\otimes R_{g}\,.
\end{equation}
\end{definition}
It is evident that they commute with all the operators $\bar U_g$, $\bar
L_{\phi(x)}$ defined by (\ref{barLU}), than also with all $\bar P_S$, and that
the operators $\bar U'_g$ are unitary. They act on $\vert \Phi \rangle= \sum_i
f(g_i, x)\otimes \vert g_i\rangle\ \equiv \sum_i f(g_ig^{-1}, x) \otimes \vert
g_ig^{-1}\rangle$ in the following way
$$
\bar L'_{\phi(x)} \vert \Phi\rangle = \hbox{$\sum_{ij}$}\Bigl[L_{\phi(g^{-1}_i
x)}\otimes P_{g_i}\Bigr]f(g_j, x)\otimes \vert g_i \rangle = \hbox{$\sum_{i}$}
\phi(g^{-1}_i x) f(g_j, x)\otimes\vert g_i\rangle \,;
$$
$$
\bar U'_g \vert \Phi\rangle = \hbox{$\sum_{i}$} \Bigl[1\otimes R_{g}\Bigr]
f(g_ig^{-1}, x) \otimes \vert g_ig^{-1}\rangle= \hbox{$\sum_{i}$} f(g_ig^{-1},
x)\otimes \vert g_i\rangle\,;
$$
or, in the abridged form,
$$
\bar L'_{\phi(x)} f(g_i, x) = \phi(g^{-1}_i x)f(g_i, x)\,,\quad \bar U'_g
f(g_i, x)  = f(g_ig^{-1}, x)\,.
$$

\begin{definition} Finally, let us define the operator
\begin{equation}
\label{barQ} \bar Q = \hbox{$\sum_i$} U_{g_i}\otimes Q_{g_i}\,,
\end{equation}
\end{definition}
which acts on $\vert \Phi \rangle= \sum_i f(g_i, x)\otimes \vert g_i\rangle
\equiv \sum_i f(g_i^{-1}, x)\otimes \vert g_i^{-1}\rangle$ as follows
$$
\bar Q \vert \Phi\rangle = \hbox{$\sum_{ij}$}\Bigl[U_{g_i}\otimes Q_{g_i}
\Bigr] f(g_j^{-1}, x)\otimes \vert g_j^{-1}\rangle
 =\hbox{$\sum_{i}$} f( g^{-1}_i,
 g^{-1}_i x) \otimes \vert g_i\rangle \,;
$$
or, in the abridged form,
$$
\bar Q f(g_i, x)  = f(g_i^{-1}, g_i^{-1}x)= U_{g_i^{-1}}f(g_i^{-1}, x)\,,\quad
g_i\in G\,.
$$
$\bar Q$ is unitary and involutory as $\bar Q^2 =1$. The second property is
evident. The first follows from the second and from the fact that
$$
\bar Q^\dag   = \hbox{$\sum_{i}$} U^\dag_{g_i}\otimes Q^\dag_{g_i}=
\hbox{$\sum_{i}$} U_{g^{-1}_i}\otimes Q_{g^{-1}_i} =  \hbox{$\sum_{i}$}
U_{g_i}\otimes Q_{g_i} = \bar Q \,.
$$
The equations
$$ \bar U'_g = \bar Q \bar U_g \bar Q\,, \quad
\bar L'_{\phi(x)} = \bar Q \bar L_{\phi(x)}\bar Q\,,\quad \bar P'_S = \bar Q
\bar P_S \bar Q
$$
can be easily verified. Eq.s (\ref{iso1}) can be isomorphically translated into
\begin{equation}
\label{iso2} \bar U'_g \,\bar L'_{\phi(x)}\bar{\;U '_g}^{-1} = \bar
L'_{\phi(gx)}\,,\quad \bar U'_g \,\bar P'_S \bar{\;U '_g}^{-1} = \bar
P'_{gS}\,.
\end{equation}

We can then introduce the commutative algebra $\bar {\cal L}^Q=\bar Q\bar {\cal
L}\bar Q$, the group  $\bar {\cal U}^Q=\bar Q\bar {\cal U}\bar Q$, the
projector family $\bar {\cal P}^Q=\bar Q\bar {\cal P}\bar Q$ and summarize the
results in the theorem
\begin{theorem}
The algebra $\bar{\cal F}= \bar {\cal U}\cup \bar {\cal L}$ generated by $\bar
{\cal U}$ and $\bar {\cal L}$ commutes with the algebra $\bar{\cal F}^Q =\bar
{\cal U}^Q\cup \bar {\cal L}^Q$ generated by $\bar {\cal U}^Q= \bar Q\bar{\cal
U}\bar Q$ e $\bar {\cal L}^Q= \bar Q\bar{\cal L}\bar Q$.
\end{theorem}

From the algebraic isomorphism  between Eq.s (\ref{iso1}) and Eq.s
(\ref{iso2}), and from the fact that the property of being maximal is
algebraic and not spatial, it follows that
\begin{theorem} The commutative algebras $\bar {\cal L}$ and  $\bar{\cal L}^Q$,
formed respectively by $\bar L_{\phi(x)}$ and  $\bar L'_{\phi(x)}$, are
maximal. The same properties hold for the  projector families $\bar{\cal P}$
and $\bar{\cal P}^Q$, which are formed respectively by $\bar P_S =\bar
L_{\chi_S(x)}$ e $\bar P'_S =\bar L'_{\chi_S(x)}$.
\end{theorem}

Now, let $\bar {\cal U}\vee \bar {\cal L}$ be the algebra of von Neumann
generated by $\bar {\cal U}$ and $\bar {\cal L}$, and $\bar {\cal U}^Q\vee
\bar {\cal L}^Q$ that generated by $\bar {\cal U}^Q= \bar Q\bar{\cal U}\bar Q$
and $\bar {\cal L}^Q= \bar Q\bar{\cal L}\bar Q$. We have

\begin{theorem}
The algebra $\bar{\cal F}=\bar {\cal U}\vee \bar {\cal L}$ commutes with $
\bar{\cal F}^Q=\bar {\cal U}^Q\vee \bar {\cal L}^Q$ and, since both of them
contain $1$, we have $\bar{\cal F}= \bar{\cal F}''$, $\bar{\cal F}^Q =
(\bar{\cal F}^Q)'' = \bar{\cal F}'$.
\end{theorem}
The first part of the theorem follows from the fact that $\bar{\cal F}$ and
$\bar{\cal F}^Q$ are respectively generated by commuting elements. The second
follows from the bicommutant theorem \ref{bicomm} and from the fact that the
commutant of an algebra of von Neumann is unique.

Since every element $\bar A \in \bar{\cal F}$ is one--to--one with $\bar Q\bar
A \bar Q\in \bar{\cal F}^Q$, and vice versa, we can state the following theorem
\begin{theorem}
If $\bar A\in \bar{\cal F}$, then $\bar Q\bar A\bar Q \equiv\bar A' \in
\bar{\cal F}'$ and if $\bar A'\in \bar{\cal F}'$, then $\bar Q\bar A'\bar Q
\equiv \bar A \in \bar{\cal F}$.
\end{theorem}

The question then arises of how $\bar{\cal F}$ and $ \bar{\cal F}'$ relate to
the algebra of all bounded operators of $\cal H$;  in particular, of whether
$\bar{\cal F}\vee \bar{\cal F}'$ exhausts the totality of bounded operators of
$\cal H$. To answer this, we need to verify whether $\bar{\cal F}$ is a factor
or if $\bar{\cal F}$, $\bar{\cal F}'$ are coupled factors.

\subsection{The coupled factors $\bar {\cal F}$, $\bar {\cal F}'$}
From the equations in the abridged form
\begin{eqnarray}
[\bar L_{\phi(x)}, A] f(g_i, x) &=& \hbox{$\sum_j$} [L_{\phi(x)}, A_{g_i}^{g_j}] f(g_j, x)\,;\nonumber\\
\nonumber\\
\bar U_g A \bar U_{g^{-1}} f(g_i, x) &=& \bar U_g A \,[U_g^{-1}f(g^{-1}g_i,x)]
= \bar U_g[\hbox{$\sum_j$} A_{g_i}^{g_j} U_{g^{-1}}f(g^{-1}g_j, x)]=\nonumber\\
& & \hbox{$\sum_j$}  U_g A_{gg_i}^{g_j} U_{g^{-1}}
f(g^{-1}g_j, x)=\hbox{$\sum_j$}U_g A_{gg_i}^{g g_j}U_g^{-1} f(g_j, x)\,;\nonumber\\
\nonumber\\
\bar Q A \bar Q f(g_i, x) &=& \bar Q A U_{g_i^{-1}} f(g_i^{-1}, x) = \bar Q
\hbox{$\sum_j$} [A_{g_i}^{g_j}  U_{g_j^{-1}} f(g_j^{-1}, x)]=\nonumber
\\ & & \bar Q \hbox{$\sum_j$}[A_{g_i}^{g_j^{-1}} U_{g_j}f(g_j,x)] = \hbox{$\sum_j$} U_{g_i^{-1}}
A_{g_i^{-1}}^{g_j^{-1}} U_{g_j}f(g_j, x)\,;\nonumber
\end{eqnarray}
we obtain the equivalence relations
\begin{equation}
\label{simAprime} [\bar L_{\phi(x)}, A] \sim  \dvert [L_{\phi(x)},
A_{g_i}^{g_j}]\,\drangle\,,\quad \bar U_g A \bar U_g^{-1} \sim \dvert  U_g
A_{gg_i}^{g g_j}U_g^{-1}\,\drangle\,,
\end{equation}
\begin{equation}
\label{simQ} \bar Q A \bar Q \sim \dvert U_{g_i^{-1}}
A_{g_i^{-1}}^{g_j^{-1}}U_{g_j}\,\drangle\,,
\end{equation}

Let us pose now the problem of determining the form of operators $A$ in case
that the equality $A=\bar A' \in \bar {\cal F}'$ holds.

Clearly, since every operator of this sort commutes with all operators of
$\bar {\cal F}$, the equations $[\bar L_{\phi(x)}, \bar A'] = 0$ and $\bar U_g
\bar A' \bar U_{g^{-1}} = \bar A'$ must hold and consequently, because of Eq.s
(\ref{simAprime}), also the equations
$$[L_{\phi(x)}, A_{g_i}^{g_j}]=0\,,\quad A_{g_i}^{g_j} =
U_g A_{gg_i}^{gg_j}U_g^{-1}\,.$$ From the first of these, since $\bar {\cal
L}$ is maximal, we deduce $A_{g_i}^{g_j} = L_{\alpha(g_i,\, g_j;\, x)}$ for
some collection of bounded measurable functions $\alpha(g_i, \,g_j; x)$. From
the second, we deduce that for every $g\in G$ we must have
$$L_{\alpha(g_i,\, g_j; \,x)} = U_g L_{\alpha(gg_i,\, gg_j; \,x)}U_g^{-1} =
L_{\alpha(gg_i,\, gg_j; \,g x)}\,.$$ This means that $\alpha(gg_i, gg_j; g x)$
does not actually depend on $g$. We can then pose $g = g_i^{-1}$ in each of
these functions and replace $\alpha(1, g_i^{-1} g_j; g^{-1}_ix)$ with
$\alpha(g_i^{-1} g_j; g^{-1}_i x)$. Thus, we can write
\begin{equation}
\label{barAprime} A_{g_i}^{g_j} = L_{\alpha(g_i^{-1} g_j;\, g^{-1}_i x)}\,.
\end{equation}
Hence, the general form of the operators of $\bar{\cal F}'$ is
\begin{equation}
\label{simAprime2} \bar A' \sim \dvert L_{\alpha(g_i^{-1} g_j;\,
g^{-1}_ix)}\,\drangle\,,
\end{equation}
and its action on a state $\vert \Phi \rangle \sim \dvert f(g_i, x)\,\drangle$
is
$$\bar A' \vert \Phi \rangle\sim \dvert\hbox{$\sum_j$}\,\alpha(g_i^{-1} g_j;\,
g^{-1}_ix)\, f(g_j, x)\,\drangle\,.$$

Let us pose also the problem of determining the form of the operators $A$ when
$A=\bar A \in \bar {\cal F}$. In this case, we can exploit the fact that $\bar
Q \bar A \bar Q = \bar A' \in \bar {\cal F}'$. It will be therefore sufficient
to find the abridged form of $\bar Q \bar A' \bar Q$.

From Eq.s (\ref{simQ}) and (\ref{barAprime}) we obtain
$$
\bar Q \bar A' \bar Q \sim \dvert U_{g_i^{-1}} L_{\alpha(g_i g_j^{-1}; g_i
x)}U_{g_j} \drangle= \dvert L_{\alpha(g_i g_j^{-1}; x)}
U_{g_i^{-1}}U_{g_j}\,\drangle\,.
$$
Hence, since $U_gU_g'= U_{g'g}$, we have
\begin{eqnarray}
\label{simA} \bar A \sim \dvert L_{\alpha(g_i g_j^{-1};\,
x)}U_{g_j\,g_i^{-1}}\,\drangle\,,
\end{eqnarray}
which acts on $\vert \Phi \rangle$ as follows
$$\bar A \vert \Phi \rangle\sim \dvert\hbox{$\sum_j$}\,
\alpha(g_i g^{-1}_j;\,x) f(g_j, g_j g_i ^{-1} x)\,\drangle\,.$$

We therefore see that a same bounded measurable function $\alpha(g_i;\,x)$ is
one--to--one with both the bounded operator $\bar A\in \bar {\cal F}$ and the
bounded operator $\bar A'\in \bar {\cal F}'$. Since the totality of such
functions is dense in $\cal H$, we can state the following theorem

\begin{theorem}
The dense set of vectors $\vert \Phi \rangle \sim \,\dvert\alpha(g_i;\,
x)\,\drangle \in\cal H$ is one--to--one with the dense set of operators $\bar
A \sim \dvert L_{\alpha(g_i g_j^{-1};\, x)}U_{g_j}\,\drangle \in \bar {\cal
F}$ on one side and the dense set operators $\bar A' \sim \dvert
L_{\alpha(g_i^{-1} g_j;\, g_i^{-1}x)}\,\drangle \in \bar {\cal F}'$ on the
other side.
\end{theorem}

Let us prove now the following structural theorems
\begin{theorem}
From the conditions that $G$ be free and ergodic on $X$, follow the equalities
$$\bar {\cal F}\wedge
\bar {\cal F}'= \{\alpha I\}\,,\quad \bar {\cal F}\vee \bar {\cal F}'= {\cal
I}\,,$$ where ${\cal I}$ is the set of all bounded operators of $\cal H$.
\end{theorem}
Were it not so, there would be an operator
$$\bar A \sim \dvert L_{\alpha(g_i g_j^{-1};\, x)}U_{g_jg_i^{-1}}\,\drangle =
\dvert L_{\alpha(g_i^{-1} g_j;\, g_i^{-1}x)}\,\drangle\,.$$ Because of theorem
\ref{free}, this is possible only if either $g_j\neq g_i$ and $$\alpha(g_i
g_j^{-1};\, x)=\alpha(g_i^{-1} g_j;\, g_i^{-1}x) =0\,\, \hbox{a.e.}$$ or
$g_j=g_i$ and $\alpha(1;\, x)=\alpha(1;\, g_i x)$, a.e. The first case implies
$A=I$, the second, because of corollary \ref{fcost}, implies $\alpha(1;\,
x)=\alpha$ (costante) a.e., hence $A=\alpha I$, because of the ergodicity
theorem \ref{ergodic}. In this way, we find the first inequality stated by the
theorem. The second follows immediately as $\{\alpha I\}'={\cal I}$.

In other terms, $\bar{\cal F}$ and $\bar{\cal F}'$ are coupled  factors. They
can be interpreted as the operator algebras of two systems, which are
physically independent, non spatially isomorphic but quantum--mechanically
coupled by the algebraic isomorphism $\bar Q\bar{\cal F}\bar Q = \bar{\cal
F}'$.

\subsection{The classification of factors}
Studying the possible coupling of two parts of a physical system is ultimately
equivalent to studying the spectral properties of the operators of $\bar{\cal
F}$ and $\bar{\cal F}'$, in particular of their projectors.

We can take advantage of the correspondence between states and operators
established in the previous subsection to study the algebraic properties of the
operators $\bar A$, ${\bar A}'$, as respectively defined by Eq.s (\ref{simA})
and (\ref{simAprime}). To express the fact that both of them correspond to the
same collection of functions $\alpha(g_i; x)$ we write.
\begin{eqnarray}
\label{dsimA} \bar A  \approx \bar A'  \approx  \dlangle \alpha(g_i;\,
x)\drangle\,.\end{eqnarray}

By substituting in Eq.s (\ref{proggA}) the expression of $\bar A$ in the form
(\ref{simA}) and that of $\bar B$ in a similar form with $\beta(g_i;x)$ in
place of $\alpha(g_i;x)$, we can immediately  verify the following
correspondences
\begin{eqnarray}
\label{corrisp} & & \lambda \bar A \approx  \dlangle \lambda\, \alpha(g_i;\,
x)\drangle\,;\\
& & \bar A^\dag \approx  \dlangle \alpha^*(g_i^{-1};\, g_i^{-1}x)\drangle\,;\nonumber \\
& & \bar A + \bar B \approx  \dlangle \alpha(g_i;\, x)+ \beta(g_i;\, x)\drangle\,;\nonumber \\
& &  \bar A\bar B \approx  \dlangle \hbox{$\sum_j $}\alpha(g_j;\,
x)\beta(g_j^{-1}g_i;\, g_j^{-1}x)\drangle\,.\nonumber
\end{eqnarray}
In particular, we have
\begin{eqnarray}
\label{AdagA} & & \bar A^\dag \bar A\approx  \dlangle \hbox{$\sum_j
$}\,\alpha^*(g_j;\,
g_jx)\,\alpha(g_jg_i;\, g_jx)\drangle\,; \\
\label{AAdag} & & \bar A \bar A^\dag\approx  \dlangle \hbox{$\sum_j
$}\,\alpha(g_j;\,x)\,\alpha^*(g_i^{-1}g_j;\, g_i^{-1}x)\drangle\,.
\end{eqnarray}

If $\bar A$ is Hermitian, the equalities $\alpha(g_i;\, x)=\alpha^*(g_i^{-1};\,
g_i^{-1}x)$ must hold; in particular, $\alpha(1;\, x)$ must be real.

If $\bar P \approx \dlangle \chi(g_i;\, x)\drangle$ is a projector, from $\bar
P = \bar P^\dag$ e $\bar P^2= \bar P$ follow the identities
$$
\hbox{$\sum_j $}\chi^*(g_j;\, g_jx) \chi(g_jg_i;\, g_jx)= \chi(g_i;\,
x)\,;\,\,\chi(g_i;\, x)=\chi^*(g_i^{-1};\, g_i^{-1}x)\,.
$$
For $g_i=1$ we have
$$
\hbox{$\sum_j $}\vert \chi(g_j;\, g_jx)\vert^2 = \hbox{$\sum_j $}\vert
\chi(g_j;\, x)\vert^2= \chi(1;\, x)\,.
$$
Therefore, if $ \chi(1;\, x)=0$ a.e., $\bar P = 0$. From the equality $\chi(1;
x) = \chi(1; x)^2$ we deduce that $\chi(1; x)$ is a box function. There exists
therefore a measurable set $S_{\bar P}$, which will be called a {\em spectral
set} of $\bar P$, on which $\chi(1; x)=1$ if $x\in S_{\bar P}$, otherwise
$\chi(1; x)=0$. Consequently, we have
\begin{equation}
\label{muP} \int_X \chi(1; x)\,d\mu(x)= \int_{S_{\bar P}} d\mu(x)=\mu(S_{\bar
P})\,,
\end{equation}
which vanishes if and only if $\bar P=0$.

If $\bar U \approx \dlangle \nu(g_i;\, x)\drangle$ is partially isometric,
then $\bar U^\dag \bar U = \bar P_1$ e $\bar U\bar U^\dag = \bar P_2$ are
projectors and from Eq.s (\ref{AdagA}), (\ref{AAdag}), with $g_i=1$, we obtain
\begin{equation}
\label{UdagU} \hbox{$\sum_i $}\vert \nu(g_i;\,g_ix)\vert^2 = \chi_1(1;
x)\,;\quad \hbox{$\sum_i $}\vert \nu(g_i; x)\vert^2 = \chi_2(1; x)\,.\\
\end{equation}
By integrating both members of these and using the equality
$d\mu(g_ix)=d\mu(x)$ we obtain
$$
\mu(S_{\bar P_1})= \int_X \chi_1(1; x) \,d\mu(x) = \int_X \chi_2(1; x)
\,d\mu(x)= \mu(S_{\bar P_2})\,,
$$
where $S_{\bar P_1}$ and $S_{\bar P_2}$ are the spectral sets of $\bar P_1$ and
$\bar P_2$.

Let us now focus on the cases in which the projectors belong to $\bar{\cal P}$;
that is, the cases in which $\chi(g_i;\,x) = \chi_S(x)\,\delta_{g_i,1}$ {\em
a.e.} If $\bar P_{S_A}, \bar P_{S_B}\in \bar{\cal P}$ e $\bar P_{S_A}\bar
P_{S_B}=0$ we will have $\bar P_{S_A}+\bar P_{S_B} = \bar P_{S_A\cup S_B}$ and,
correspondingly, $\chi_{S_A}(x)\chi_{S_B}(x)=0$, $\chi_{S_A}(x)+ \chi_{S_B}(x)
= \chi_{S_A\cup S_B}(x)$. We will obtain therefore
$$
\int_X \chi_{S_A\cup S_B}(x)\,d\mu(x)= \mu(S_A) + \mu(S_B)\,.
$$
This additive property of measures transfers immediately to all pairs of
projectors of $\bar P_1, \bar P_2 \in \bar {\cal F}$ for which the relations
$\bar P_1\bar P_2=0$,  $\bar P_1 \sim \bar P_{S_A}$, $\bar P_2 \sim \bar
P_{S_B}$ hold.

By noting that, $\bar{\cal P}$ being maximal, every projector $\bar P\in
\bar{\cal F}$ is isometric to a projector $\bar P_S \in \bar{\cal P}$ at
least, we deduce the additive property
$$
\mu(S_{\bar P_1 + \bar P_2}) = \mu(S_{\bar P_1})+\mu(S_{\bar P_2})\,,\quad
\bar P_1\bar P_2=0
$$
does hold.

The same results can be obtained, of course, also if $\bar P'\in \bar{\cal
F}'$ for all projectors. It is then evident that $\mu(S_{\bar P})$ has all the
properties of a relative dimension $D(\bar P)$. We can then establish a
proportionality relation $D(\bar P)= c \mu(S_{\bar P})$, where $c$ is a real
positive normalization constant. We can therefore state
\begin{theorem}
\label{DeqMu} Every projector of the sort $\bar P \approx \bar P' \approx
\dlangle \chi(g_i;\, x)\drangle$ has dimension
$$D(\bar P)=D(\bar P')=c\int_X \chi(1; x)\,d\mu(x)\,,$$
where $c$ is a suitable normalization constant.
\end{theorem}
We can establish in this way a precise correspondence between the dimensions of
the projectors and the  measures of their spectral sets in all cases in which
these measures are invariant with respect to $G$. That is, for all factors of
types $I$ and $II$.

Here are some examples of discrete groups that act on $X$ so as to generate
the described types:
\begin{itemize}
\item[$I_n$.] The additive group over the integers modulo $n$ $Z_n$ on $X=|Z_n|$,
the set of $Z_n$ elements.
\item[$I_\infty$.] The additive group $Z$ over the integers on $X=|Z|$, the set of $Z$ elements.
\item[$II_1$.] The additive group $Q_1$ modulo 1 generated by  the rational
numbers $g_n=1/n$, $n=1,2,\dots$, which acts on the real numbers $x\in[0,1]$,
or on the set $|Q_1|$ itself, according to the law $x'=x+g$.
\item[$II_\infty$.] The additive group of rational numbers $Q$ that acts on the
real numbers $x\in[-\infty,+\infty]$, or on the set $|Q|$ itself, according to
the law $x'=x+g$.
\end{itemize}

\subsection{The class Trace}
If the measure of $X$ is finite, we can normalize it so as to have $\mu(X) =
1$. In this case, $\bar {\cal F}$ and $\bar {\cal F}'$ belong to types $I_n$ or
$II_1$. The type $I_n$ occurs when $X$ is a set of $n$ points, in which case
the measure of a point is $\mu(x)=1/n$. The type $II_1$ occurs when the
projectors $\bar{\cal P}$ have a continuous spectrum of finite measure.
Correspondingly, the vector $\vert \Omega \rangle = 1\otimes \vert 1\rangle$
has the finite norm
$$
\langle\Omega \vert \Omega \rangle = \int_X d\mu(x) =1\,.
$$
We can therefore establish the equations
\begin{eqnarray}
& & \bar A^\dag \bar A\vert \Omega \rangle = \hbox{$\sum_j $}\vert
\alpha^*(g_j;\,
g_jx)\vert^2 \otimes \vert g_j \rangle\,;\nonumber\\
& & \bar A \bar A^\dag\vert \Omega \rangle = \hbox{$\sum_j $}\vert
\alpha^*(g_j;\,
g_jx)\vert^2 \otimes \vert g_j \rangle\,;\nonumber\\
& & \bar P\vert \Omega \rangle = \hbox{$\sum_j $}\chi(g_j;\, g_j x) \otimes
\vert g_j \rangle\,;\nonumber
\end{eqnarray}
from which we obtain
$$
\langle \Omega\vert \bar A \bar A^\dag\vert \Omega \rangle = \langle
\Omega\vert \bar A^\dag \bar A\vert \Omega \rangle=\hbox{$\sum_j $}\int_X \vert
\alpha(g_j;\,x)\vert^2d\mu(x)\,.
$$
$$
\langle \Omega\vert\bar P\vert \Omega \rangle = \int_X \chi(1;\, x)\mu(x)=
D(\bar P)\le 1 \,.
$$
If $\bar U \in \bar{\cal F}$, is unitary, we have $\bar P \sim \bar U \bar P
\bar U^\dag$ and, in agreement with the note following Def. \ref{isometrici},
we have also $\langle\Omega\vert\bar P\vert \Omega \rangle = \langle
\Omega\vert\bar U \bar P \bar U^\dag \vert\Omega \rangle$.

Let us prove that also every self--adjoint operator $\bar A \in \bar{\cal F} $
satisfies the equation
$$\langle\Omega\vert\bar A\vert \Omega \rangle = \langle \Omega\vert\bar U
\bar A \bar U^\dag \vert\Omega \rangle\,.$$ It is indeed sufficient to
represent $A$ in the form
$$\bar A = \int_0^1 \alpha\, d\bar E(\alpha)\,, $$
where $\bar E(\alpha)\in \bar{\cal F}$ is a suitable {\em matrioska} of
projectors depending on the real parameter $\alpha$, $0\le \alpha \le 1$. That
is, a family of projectors such that $\bar E(\alpha_1)\bar E(\alpha_2) = \bar
E(\alpha_1)$ se $\alpha_1 \le \alpha_2$. Clearly, if $\bar U \in \bar{\cal F}$
is unitary, we have $\bar E(\alpha) \sim \bar U\bar E(\alpha)\bar U^\dag$ for
every value of $\alpha$. Consequently,
$$\langle\Omega\vert\bar E(\alpha)\vert \Omega \rangle = \langle
\Omega\vert\bar U \bar E(\alpha) \bar U^\dag \vert\Omega \rangle =\int_0^1
d\bar E(\alpha)= \mu(\alpha)\,,$$ where $\mu(\alpha)$ is the dimension of the
projector $\bar E(\alpha)$. But then, the mean value can be expressed as
$$\langle \Omega\vert\bar U \bar A \bar U^\dag \vert\Omega \rangle =
\int_0^1 \alpha\, d\mu(\alpha)\,,$$ which is clearly independent of $\bar U$.
From all of this, it follows that determining the mean value in $\vert \Omega
\rangle$ is equivalent to computing the trace:
$$\langle \Omega\vert \bar A \bar B \dots \vert\Omega \rangle  \equiv \hbox{Tr}[\bar
A \bar B \dots]\,.$$

\subsection{The density matrix}
All projectors of a factor of type $I_\infty$, and therefore all of its
self--adjoint projectors, have an infinite discrete spectrum and all those of
type $II_\infty$ have an infinite continuous spectrum. These are just the
cases of interest for quantum mechanics. It is clear, however, that the
considerations carried out in the previous subsection do not apply to these
types.

The difficulty lies in the fact that the mean values of operators do not exist.
Otherwise, the norm of any hypothetical state $\vert\Omega \rangle=1\otimes
\vert 1\rangle$ would be infinite.

From a physical standpoint, the non--existence of the trace is related to the
fact that the probability distributions of the eigenstates of an observable
with an infinite spectrum cannot be uniform, otherwise we should allow a sense
to infinite sets of vanishing probabilities. Types $I_\infty$ and $II_\infty$
can host the representations of physical systems for which the mean values of
observable quantities, no matter whether discrete or continuous, are evaluated
over non uniform probability distributions (provided that the eigenvalue
spectra do not have infinite degeneration). Of this sort are for instance
finite quantum--mechanical systems in thermodynamic equilibrium. In this case,
the non--uniformity of the probability distributions is a consequence of the
exponentially decreasing profile of Gibbs' distribution function.

With maximum generality, the means of physical quantities can be calculated as
mean values of operators over states represented by vectors of the sort
$$
\vert\Omega \rangle = \sum_i f(g_i, x)\otimes \vert g_i\rangle\,,\quad\langle
\Omega\vert \Omega \rangle = \sum_{i} \int_X \vert f(g_i, x)\vert^2 d\mu(x)
=1\,.
$$
By setting
$$
w_{g_i}^{g_j}(x) = f^*(g_i; g_ix) f(g_j; g_jx)\,;\,\,w(x) = \hbox{$\sum_{i}$}
w_{g_i}^{g_i}(x) = \hbox{$\sum_{i}$} \vert f(g_i, x)\vert^2 \,,
$$
we obtain for $\bar L_{\phi(x)}\in \bar{\cal L}$
$$
\langle \Omega\vert\bar L_{\phi(x)}\vert\Omega \rangle = \int_X w(x)\,\phi(x)\,
d\mu(x) \,.
$$

This expression is equivalent to evaluating the mean value of $\phi(x)$ over a
probability density $w(x)$.

For a generic bounded operator $\bar A \sim \dvert L_\alpha(g_ig_j^{-1};x)
U_{g_ig_j^{-1}}\drangle \in \bar{\cal F}$ we obtain
$$
\langle \Omega\vert \bar A \vert\Omega \rangle = \hbox{$\sum_{i,j}$}\int_X
w_{g_i}^{g_j}(x)\, \alpha(g_ig_j^{-1}; g_ix)\, d\mu(x) \,,
$$
We can interpret $w_{g_i}^{g_j}(x)$ as the matrix elements of a state--density
operator $W$ of $\cal H$.

In the case of type $I_\infty$, the space $X$ contains a countable infinity of
points $x_1, x_2,\dots$ with same $\mu(x_k)$. Since $\mu(X)=\infty$, it is not
restrictive to assume  $\mu(x_k)=1$. In this way, the integration becomes a
summation over the indices $k$ and the averages indicated above became
respectively
$$\langle \Omega\vert\bar L_{\phi(x)}\vert\Omega \rangle =
\sum_{k=1}^{\infty}w(x_k)\,\phi(x_k) \,, $$
$$
\langle \Omega\vert \bar A \vert\Omega \rangle =
\sum_{i,j,k}w_{g_i}^{g_j}(x_k)\, \alpha(g_ig_j^{-1}; g_i x_k) \,.
$$

\section{Factors of Type $III$}
In the previous section, we have proved that for factors of types $I$ and $II$
there is a substantial equivalence between the dimensions of the projectors
and the measures of their spectral sets.

In the proof, the invariance axiom (x) introduced in subsection \ref{assiomi}
was used. This ensures that the equality $\mu(gS)= \mu(S)$ holds for all
measurable sets $S\subset X$ and all elements $g \in G$. This means that, in
the framework of discrete systems here considered, possible examples of
factors of type $III$ must be searched for under the condition that axiom (x)
does not hold. That is, when the inequality $\mu(gS)\neq \mu(S)$ holds for some
$S\subset X$ and some $g \in G$. Let us here provide two examples of such
groups, the first of which has $\mu(X)=\infty$ and the second $\mu(X)<\infty$.

\subsection{Examples of groups with non--invariant measures}
\begin{itemize}
\item[1.] Let $X$ be the set of real numbers equipped with standard Lebesgue measures.
Hence we have $\mu(X)=\infty$. Let $\bar G$ be the group of affine
transformations defined by the equations
$$
x'=g_{\rho, \sigma}x= \rho \,x+\sigma\,,\quad x\in X\,,
$$
and $G$ the subgroup of $\bar G$ with rational $\rho$ and $\sigma$. Clearly,
$G$ is discrete free and ergodic and, if $S\subset X$ is measurable, we have
$$\mu(g_{\rho, \sigma}S) = \rho\,\mu(S)\,.$$ Therefore the measure is not
invariant.

\item[2.] Let $X$ the set of complex numbers $z$ with $|z| =1$. Assume as
measurable sets those formed by the arcs of the unitary circle $|z|=1$. Then
$\mu(X)=2\pi$.

Let now $\bar G$ be the group of conformal maps
$$
z' =g_{\rho, \sigma}z= e^{i2\pi \rho} \frac{z+\sigma}{1+
z\sigma^*}\,,\quad\rho\in[0,1)\,;\,\,|\sigma|<1\,.
$$
We can easily verify that $|z'|=1$. In other terms $\bar G$ maps the unitary
circle on itself. Let $G$ be the subgroup of $\bar G$ generated by $g_{0, 1/2}$
and $g_{\rho, 0}$ with rational $\rho$. Clearly, $G$ is infinite and discrete.
We can immediately verify that it is free and ergodic: Indeed,  ergodic is
already the subgroup $G_0$ formed by the sole elements of $g_{\rho, 0}$. Now,
indicating by $S_{0,\pi/2}$ the arc delimited by the complex vectors $z=1$ and
$z=i$, we have $\mu(S_{0,\pi/2})= \pi/2$. But we have $\mu(g_{0, 1/2}
S_{0,\pi/2}) = \arctan 3/4$ as $g_{0, 1/2}$ sends 1 into 1 but $i$ into
$(4+3i)/5$. Thus, in general, the measures of the arcs are not invariant under
$G$.
\end{itemize}

\subsection{The theorem of Radon-Nikodym}
We can define unitary transformations that do not leave invariant the spectral
sets of projectors by using a theorem of Lebesgue--Radon--Nikodym adapted by
von Neumann to deal with sets that are measurable in the generalized sense
defined in subsection \ref{assiomi}. The theorem, which we report without proof
\cite{NEU3}, states the following:
\begin{theorem}
Let $\mu_1(S)$ and $\mu_2(S)$ different measures of a measurable subset
$S\subset X$. Then, there exists a positive measurable function $\kappa(x)$,
which is in general defined up to a set of measure zero, such that for all
measurable functions $f(x)$ the following equations
$$ \int_X f(x)\,d\mu_2(x) = \int_X f(x)\,\kappa(x)\,d\mu_1 (x)$$ hold.
We can therefore represent this function as
$$ \kappa(x) = \frac{d\mu_2 (x)}{d\mu_1 (x)}\,.$$
\end{theorem}

We can apply this theorem to the case in which the diversity of the measure
depends upon the non--invariance under $G$, so as to give a sense to the
expression
$$ \frac{d\mu(g_2x)}{d\mu(g_1 x)}= \frac{d\mu(g_1^{-1} x)}{d\mu(g_2^{-1} x)}.$$

\subsection{The algebras of systems with non--invariant measures}
The condition that the operators $U_g$, which represent the elements $g\in G$,
be unitary imposes the following generalization of the operators introduced in
Subsec. \ref{algdiscr}
$$ L_\phi(x) f(x)= \phi(x) f(x)\,,\quad  U_g f(x)=
f(gx) \Bigr[\frac{d\mu(gx)}{d\mu(x)}\Bigl]^{1/2}\,.
$$
We can immediately verify that $U_g$ is unitary. In a similar way, for the
operators of ${\cal H}={\cal H}_X\otimes {\cal H}_G$, we can define
$$
\bar L_\phi(x) f(g_i, x)= \phi(x) f(g_i, x)\,;\,\, \bar U_g f(g_i, x)= f(gg_i,
gx)\Bigr[\frac{d\mu(gx)}{d\mu(x)}\Bigl]^{1/2}\,;
$$
$$
\bar Q  f(g_i, x)= f(g_i^{-1},
g_i^{-1}x)\Bigr[\frac{d\mu(g_i^{-1}x)}{d\mu(x)}\Bigl]^{1/2}\,;
$$
$$
\bar L'_\phi(x) f(g_i, x)= \phi(g_i^{-1}x) f(g_i, x)\,;\quad  \bar U'_g f(g_i,
x)= f(g_ig^{-1},x) \,.
$$
It is easy to verify that $\bar P_S=\bar L_{\chi_S(x)}$ e $\bar P_S'=\bar
L'_{\chi_S(x)}$ are projectors, that $\bar U_g$, $\bar Q$ e $\bar U'_g$ are
unitary and that the equalities
$$Q^2=Q\,,\quad \bar Q \bar L_\phi(x) \bar Q =
\bar L'_\phi(x)\,,\quad\bar Q \bar U_g \bar Q =\bar U'_g$$ hold.

We can therefore proceed to the construction of all the representations of
operators $\bar A \in {\cal F}$ and  $\bar A' \in {\cal F}'$ exactly as they
were in the case of invariant--measure systems. We proceed up to the point at
which, for the spectral sets $S_{\bar P}$ and $S_{\bar P_g}$ of two projectors
$\bar P, \bar P_g$ linked by the equation $\bar P_g = \bar U_g \bar P \bar
U_g^\dag$, we find $S_{\bar P_g} = g S_{\bar P}$ a.e., and finally $D(\bar P) =
c\mu(S_{\bar P})$. Since $\bar P$ and $\bar P_g$ are isometric, the equation
$D(\bar P) = D(\bar P_{g})$ must hold.

If we pose the condition that the equality holds for all $g\in G$ and all
measurable sets $S\subset X$ with $c\neq 0, \infty$, then for all projectors of
finite spectral measure the equations $D(\bar P) = c \mu(S_{\bar P}) < \infty$
must hold. In this case, we can assume $\mu(S) = D(\bar P_{S})$ and use the
dimensions of projectors as invariant measures of their spectral sets. On this
basis, we can establish a direct relation between the measurability properties
of spectral sets and the dimensional properties of types $I$ and $II$.

However, if at least for one element $g\in G$ and one subset $S\subset X$ we
have $\mu(S_{\bar P})\neq \mu(S_{\bar P'})$, we must conclude that the relation
between dimensions and measures is possible only if either $c=0$ or $c=\infty$.
This means that the factor is of type $III$.

We can summarize these results in the theorem:
\begin{theorem}
The classification of factor types is substantially based on the
proportionality relation $D(\bar P) = c \mu(S_{\bar P})$ between the dimensions
of projectors $\bar P$ and the measures of spectral sets $S_{\bar P}$ (as
established by theorem \ref{DeqMu}), and on the invariance of this relation
under the action of unitary automorphisms $\bar U_g\bar P\bar U_g^\dag$
internal to the factor. If the measure of a spectral set is not invariant for
one of these automorphisms, the equality is possible only for $c=0$ or for
$c=\infty$, and therefore the factor is of type $III$.
\end{theorem}
This theorem does not state that the spectral sets of the projectors of type
$III$ are not measurable, but that the dependence of the measure from certain
unitary or isometric transformations makes it impossible to assign a same
finite dimension to isometric projectors. From a physical standpoint, we can
say that type $III$ is characteristic of physical systems in which some
observables can undergo scale transformation under the action of certain
physical operations.

\section{Mean values and measurement processes}
In all generality, the mean values of bounded operators of the three types can
be evaluated over normalized vectors
$$
\vert\Omega \rangle = \sum_i f(g_i, x)\otimes \vert g_i\rangle\,,\quad\langle
\Omega\vert \Omega \rangle = \sum_{i} \int_X \vert f(g_i, x)\vert^2 d\mu(x)
=1\,.
$$
For a generic bounded operator $A=\dvert A_{g_i}^{g_j}\drangle$ of $\cal H$ we
have
$$
\langle \Omega\vert A \vert\Omega \rangle = \hbox{$\sum_{i,j}$} \int_X f^*(g_i;
x) f(g_j; x) A_{g_i}^{g_j}\, d\mu(x) \,.
$$
For an operator $\bar A \sim \dvert L_{\alpha(g_i g_j^{-1};\,
x)}U_{g_j\,g_i^{-1}}\drangle \in \bar {\cal F}$ we find
$$
\langle \Omega\vert \bar A \vert\Omega \rangle = \hbox{$\sum_{i,j}$} \int_X
f^*(g_i; x) f(g_j; g_jg_i^{-1}x)\, \alpha(g_i g_j^{-1};
x)\Bigr[\frac{d\mu(g_jg_i^{-1}x)}{d\mu(x)}\Bigl]^{1/2} d\mu(x) \,,
$$
and for $\bar A' = \bar Q \bar A \bar Q \sim \dvert L_{\alpha(g_i^{-1} g_j;
g_i^{-1} x)}\drangle\in \bar {\cal F}'$
$$
\langle \Omega\vert \bar A' \vert\Omega \rangle = \hbox{$\sum_{i,j}$} \int_X
f^*(g_i; x) f(g_j; x)\, \alpha(g_i^{-1} g_j; g_i^{-1} x)\, d\mu(x) \,.
$$

Therefore, in general, we have $\langle \Omega\vert \bar A \vert\Omega \rangle
\neq \langle \Omega\vert \bar A' \vert\Omega \rangle$. It is therefore evident
that $\bar A $ e $\bar A'$, although algebraically isomorphic, are spatially
different.

However, if $\alpha(g_i; x) = \alpha(1; x) \,\delta_{g_i, 1}$, the expectation
values of ${\bar A}$ e ${\bar A'}$  coincide over all states. In this case, we
have
$$ \langle \Omega\vert \bar A \vert\Omega \rangle = \langle \Omega\vert
\bar A' \vert\Omega \rangle=\hbox{$\sum_{i}$} \int_X \vert f(g_j; x)\vert^2\,
\alpha(1;x)\, d\mu(x) \,.
$$
By interpreting $\hbox{$\sum_{i}$}\vert f(g_j; x)\vert^2 = w(x)$ as a
probability density and $a(x)=\alpha(1;x)$ as an eigenvalue spectrum, we
conclude that $\bar A $ and $\bar A'$ possess the same mean value and the same
standard deviation. In correspondence with different functions $\alpha(1;x)$,
we will find different pairs of operators, which can be interpreted as a
correlation of observables between the two parts of a system.

These are just the conditions that are expected in a measurement process.
Conditions in which, for instance, $\bar A$ represents the observed quantity
and $\bar A'$ the measuring device. It is worth noticing that the algebraic
factorization allow us to deal equally well with the measures of observables
with discrete and with continuous spectra. We answer in this way a question
that was posed at the beginning of the paper: how to represent the correlation
between the states of an observed and an observing system when the vectors of
the pair of correlated systems cannot be represented as vectors of a
direct--product space.\footnote{At this point, we should be able to describe
the unitary entanglement--operators capable of producing the correlation of
the states of initially uncorrelated observables. We leave the solution of this
problem as an exercise for the reader.}

Actually, the subject pertains more in general to the way in which two parts of
a system correlate with each other during their interaction. Indeed, we know
that during the evolution of an isolated system, it occurs in general the
exchange of conservative quantities, which are observable quantities common to
the different parts of the system. The representation of interactions in
direct--product spaces can only account for exchanges of quantities possessing
discrete spectra of eigenvalues (quanta). The factorization of the algebra of
observables into commuting subalgebras of different types, we arrive to
represent such exchanges at the level of maximum generality.

At this point, however, we should point out that the decomposition of an
algebra into a pair of coupled factors is possible only if the system is
decomposed into two algebraically isomorphic parts, which excludes the case of
asymmetric partitions. To treat the general case, we should study the
asymmetric decompositions of the bounded operator algebra into coupled factors.
This can be done by suitably selecting, out of one coupled factor, a
subalgebra as an asymmetric factor. We will not try, however, to expand our
analysis in this direction.

\subsection*{Acknowledgments}
The Author is grateful to Kurt Lechner, Pieralberto Marchetti,  Marco Matone ,
Mario Tonin and Paolo Pasti for appreciated comments and suggestions.

\markright{R.Nobili, Bibliography}
\bibliographystyle{plain}

%\newpage

%\pagestyle{plain}

%\bibliographystyle{plain}

\end{document}